\documentclass[preprintnumbers,amsmath,amssymb,twocolumn,superscriptaddress]{revtex4}

\usepackage{graphicx}
\usepackage{subfigure}
\usepackage{dcolumn}
\usepackage{bm}
\usepackage{booktabs}
\usepackage{threeparttable}
\usepackage{multirow}
\usepackage{array}
\usepackage{makecell}
\usepackage{enumerate}
\usepackage[colorlinks,
            citecolor=blue,
            anchorcolor=red,
            menucolor=red,
            linkcolor=red,
            filecolor=red,
            runcolor=red,
            urlcolor=blue,
            frenchlinks=red]{hyperref}

\newcommand{\PreserveBackslash}[1]{\let\temp=\\#1\let\\=\temp}
\newcolumntype{C}[1]{>{\PreserveBackslash\centering}p{#1}}
\newcolumntype{R}[1]{>{\PreserveBackslash\raggedleft}p{#1}}
\newcolumntype{L}[1]{>{\PreserveBackslash\raggedright}p{#1}}

\begin{document}

\title{Photoproduction of dileptons, photons and light vector mesons in p-p, p-Pb and Pb-Pb collisions at Large Hadron Collider energies}

\author{Zhi-Lei Ma}
\affiliation {Department of Physics, Yunnan University, Kunming 650091, China}
\affiliation {Department of Astronomy, Key Laboratory of Astroparticle Physics of Yunnan Province, Yunnan University, Kunming 650091, China}

\author{Zhun Lu}
\affiliation {School of Physics, Southeast University, Nanjing 211189, China}

\author{Jia-Qing Zhu}
\email{zhujiaqing@ynu.edu.cn}
\affiliation {Department of Physics, Yunnan University, Kunming 650091, China}

\author{Li Zhang}
\email{lizhang@ynu.edu.cn}
\affiliation {Department of Astronomy, Key Laboratory of Astroparticle Physics of Yunnan Province, Yunnan University, Kunming 650091, China}

\date{\today}

\begin{abstract}
The photoproduction of large $p_{T}$ dileptons, photons and light vector mesons in p-p, p-Pb and Pb-Pb collisions at LHC energies is studied, where the fragmentation processes and the ultra-incoherent photon channel are included.
An exact treatment is developed for photoproduction processes in heavy ion collisions, which recovers the equivalent photon approximation (EPA) in the limit $Q^{2}\rightarrow0$ and can avoid double counting effectively.
The full kinematical relations are also achieved.
We present the results as the distributions in $Q^{2}$, $p_{T}$ and $y_{r}$, the total cross sections are also estimated.
The numerical results indicate that:
the contribution of photoproduction processes is evident in the large $p_{T}$ and $y_{r}$ regions, and starts to play a fundamental role in p-Pb collisions.
The ultra-incoherent photon emission is a important channel of photoproduction processes, which can provide the meaningful contributions.
EPA is only applicable in small $y$ and $Q^{2}$ domains, and is very sensitive to the values of $y_{\mathrm{max}}$ and $Q^{2}_{\mathrm{max}}$.
The EPA errors appear when $y>0.29$ and $Q^{2}>0.1~\mathrm{GeV}^{2}$, and are rather serious in p-Pb and Pb-Pb collisions.
When dealing with widely kinematical regions, the exact treatment needs to be adopted.
\end{abstract}


\maketitle

\section{INTRODUCTION}

Photoproduction processes are such reactions that a photon from the projectile interacts with the hadronic component of the target.
The traditional method for studying these types of interactions is equivalent photon approximation (EPA), which can be traced back to early works by Fermi, Weizs\"{a}cker and Williams, and Landau and Lifshitz~\cite{Fermi:1924tc, vonWeizsacker:1934nji, Williams:1934ad, Sov.Phys._6_244}.
The central idea of EPA is that the electromagnetic field of a fast charged particle can be interpreted as an equivalent flux of photons distributed with some density $n(\omega)$ on a frequency spectrum~\cite{Phys.Rev._51_1037, Dalitz:1957dd, Nucl.Phys._23_1295}.
Therefore, the cross section can be approximated by the convolution of the photon flux with the relevant real photoproduction cross section.
The photon flux is the very important function which significantly decides the accuracy of photoproduction processes.
Since the convenience and simplicity of EPA, photoproduction processes have been investigated both experimentally and theoretically.
First of all, photoproduction of dileptons, photons and low-mass vector mesons $(\rho,~\omega~\textrm{and}~\phi)$ can provide valuable information on the hot and dense state of strongly interacting matter, and low-mass vector mesons can also be used to test the non-perturbative regime of QCD~\cite{Incani:2012ng, Santos:2014vwa, Guzey:2018bay, Fontannaz:2001ek, Manohar:2016nzj, Mariotto:2013qsa, Ma:2018zzq}.
Secondly, inclusive photonuclear processes are of particular interests for the study of small-x parton densities, while dijet~\cite{Vogt:2004yr}, heavy flavor~\cite{Klein:2002wm} and quarkonia photoproduction can be applied to extract small-$x$ gluon densities in protons and nuclei~\cite{Salgado:2011wc}.
Thirdly, exclusive production of heavy vector mesons $(J/\Psi, \Upsilon)$ offers a useful approach to constrain the small-x nuclear gluon density and provides a rather direct measurement of nuclear shadowing~\cite{DeGruttola:2014kta}.
Finally, photoproduction mechanism plays a fundamental role in the $\emph{ep}$ deep inelastic scattering at the Hadron Electron Ring Accelerators \cite{Butterworth:1996zw}, and is also an important part of current experimental efforts at the Large Hadron Collider (LHC) \cite{Acharya:2019vlb}.
Besides, it is the dominant channel in ultra-peripheral collisions~\cite{Baltz:2007kq, Nucl.Rev.Part.Sci_55_271, Baltz:2002pp, Djuvsland:2010qs, Klein:1999qj}.
Because of these interesting features, photoproduction processes are recognized as a remarkable tool to improve our understanding of strong interactions at high energy regime.

Although the tremendous successes have been achieved, the discussion about the accuracy of EPA and its applicability range is still inadequate.
EPA is usually adopted to processes which are actually not applicable, and a number of imprecise statements and some widely used equivalent photon spectra are obtained beyond the EPA validity range~\cite{Drees:1989vq, Drees:1988pp, Frixione:1993yw, Zhu:2015via, Zhu:2015qoz, Fu:2011zzm, Fu:2011zzf, Chin.Phys.C_36_721, Yu:2015kva, Yu:2017rfi, Yu:2017pot, Nystrand:2004vn, Nystrand:2006gi, Kniehl:2001tk, Kniehl:1990iv, sp}.
Especially in the case of heavy-ion collisions at LHC energies, the validity of EPA is crucial to the accuracy of photoproduction processes, since its influence is enlarged by the high photon flux.
The equivalent photon flux scales as the square of nuclear charge $Z^{2}$, which is a large enhancement factor for the cross section.
Thus heavy ions have a considerable flux advantage over proton, especially at the LHC energies the intense heavy-ion beam represents a prolific source of quasireal photons.
For these reasons, we consider that it is meaningful and necessary to derive in details EPA in heavy-ion collisions
and to discuss important errors and inaccuracies encountered in its application.


On the other hand, there are two types of photon emission mechanisms in high energy heavy-ion collisions: coherent-photon emission (coh.) and incoherent-photon emission (incoh.).
In the first type, the virtual photons are radiated coherently by the whole nucleus which remains intact after photons emitted.
In the second type, the virtual photons are emitted incoherently by the individual constituents (protons or even quarks) inside nucleus, and as a weakly bound system nucleus will dissociate after photons emitted.
For convenience, in the second type we further denote the process in which photons emitted from protons inside nucleus as ordinary-incoherent photon emission (OIC.), and denote that from quarks inside nucleus as ultra-incoherent photon emission (UIC).
When different photon emission mechanisms are considered simultaneously, we have to weight these different contributions for avoiding double counting.
But in fact, this serious trouble is encountered in most works and caused the large fictitious contributions~\cite{Zhu:2015qoz, Fu:2011zzm, Fu:2011zzf, Chin.Phys.C_36_721, Yu:2015kva, Yu:2017rfi, Yu:2017pot}.

Furthermore, there are a lot of studies for these photon emission processes, and the ultra-incoherent photon emission mechanism has been used in the two-photon processes~\cite{Drees:1994zx, Ohnemus:1993qw}.
However, the application of this mechanism from the individual quarks, to our knowledge, is insufficient in photoproduction processes.
Authors in Ref.~\cite{Dyndal:2019ylt} calculated the cross sections for photon-induced dileptons production in p-Pb collisions at LHC, and then they used these processes to probe the photonic content of the proton.
Authors in Ref.~\cite{Aurenche:2011wu} calculated the inclusive production of prompt photons in DIS, which involves direct, fragmentation and resolved contributions.
And they compared the theoretical predictions with H1 and ZEUS data.
In Ref.~\cite{Klein:2016yzr}, Klein and Nystrand presented a Monte Carlo simulation program, STARTlight, which calculated the cross sections for a variety of UPC final states, where the light vector mesons photoproduction are discussed.
In Ref.~\cite{Dittmaier:2009cr}, Dittmaier and Huber studied the dileptons production processes involving photons in the initial state, where the electroweak corrections are included.
There are also a lot of other works for the photoproduction of theses final states.
However, the photon emission types in all of theses above works are coherent.

According to the purposes discussed above, in the present work, we investigate the photoproduction of large $p_{T}$ photons, dileptons and light vector mesons in p-p, p-Pb and Pb-Pb collisions at LHC energies.
An exact treatment is developed which recovers EPA when the virtuality of photon $Q^{2}\rightarrow0$ and can avoid double counting effectively, where the effect of magnetic form factor are included.
The relevant kinematical relations matched with the exact treatment are also achieved.
We present the comparisons between the EPA results and the exact ones as the distributions in $Q^{2}$, $p_{T}$ and $y_{r}$, the total cross sections are also estimated.

The rest of the paper is organized as follows.
In Section.~\ref{sec:GeneralF_ET}, we present the formalism of exact treatment for the photoproduction of large $p_{T}$ dileptons, photons and light vector mesons in p-p, p-Pb and Pb-Pb collisions, where the direct, resolved and fragmentation contributions are involved.
Based on the method of Martin and Ryskin, the coherent, ordinary-incoherent and ultra-incoherent contributions are considered simultaneously.
In Section.~\ref{EPASp}, we switch the formulae of exact treatment to the approximate ones of EPA by taking $Q^{2}\rightarrow0$, and discuss the several widely applied equivalent photon fluxes.
In Section.~\ref{NUMERICAL RESULTS}, we present the numerical results of the distributions in $Q^{2}$, $p_{T}$ and $y_{r}$, and the total cross sections at LHC energies.
We summarize the paper in Section. \ref{Sum.con}.

\section{General formalism of exact treatment}
\label{sec:GeneralF_ET}

\begin{figure}[htpb]
\centering
\includegraphics[width=0.7\columnwidth]{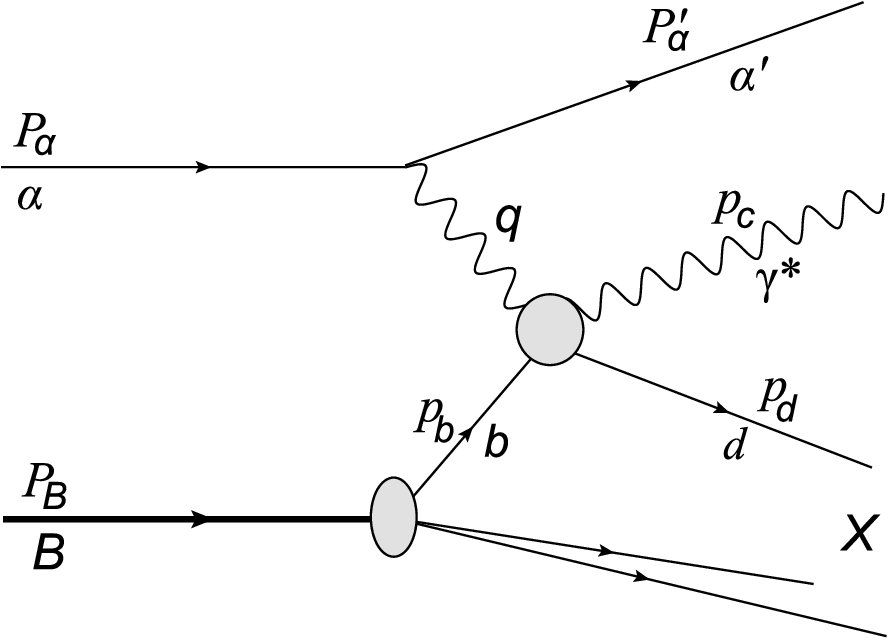}
\caption{The general photoproduction processes, the virtual photon emitted from the projectile $\alpha$ interacts with parton $b$ in nucleus $B$.
$X$ is the sum of residue of $B$ after scattering with photon. $\alpha$ can be the nucleus or its  parton (protons or quarks).}
\label{fig:gen_coll}
\end{figure}

A consistent analysis of the terms neglected in going from the accurate expression of diagram of Fig.~\ref{fig:gen_coll} to the EPA one permits in a natural manner to estimate the the applicability range of EPA and its accuracy.
As a generalization of Leptoproduction framework, the exact treatment consist of two important parts.
Firstly, the photon radiated from the projectile is off mass shell and no longer transversely polarized, thus we can expand the density of this virtual photon by using the transverse and longitudinal operators.
Secondly, the square of the electric form factor $F^{2}_{1}(Q^{2})$ is used as the probability or weighting factor (WF) to distinguish the contributions from the different photon emission processes, and thus the double counting can be avoided.

\subsection{The accurate expression of cross section for the general process $\alpha B\rightarrow\alpha \gamma^{*}X$}
\label{subsec:Acu.ex.}
The general form of cross section for the process $\alpha + B\rightarrow \alpha + \gamma^{*} +X$ in Fig. 1 is
\begin{eqnarray}\label{Gen.DCS}
&&d\sigma(\alpha+B\rightarrow \alpha+\gamma^{*}+X)\nonumber\\
&=&\sum_{b}\int dx_{b}f_{b/B}(x_{b},\mu_{b}^{2})d\sigma(\alpha+b\rightarrow \alpha+\gamma^{*}+d),
\end{eqnarray}
where $x_{b}=p_{b}/p_{B}$ is the parton's momentum fraction, $f_{b/B}(x_{b},\mu_{b}^{2})$ is the parton distribution function of massless parton $b$ in nucleus $B$,
\begin{eqnarray}\label{fp}
f_{i}^{A}(x_{i},\mu_{i}^{2})=R_{i}^{A}(x_{i},\mu_{i}^{2})\left[Zp_{i}(x_{i},\mu_{i}^{2})+Nn_{i}(x_{i},\mu_{i}^{2})\right],
\end{eqnarray}
where the factorized scale is chosen as $\mu_{i}=\sqrt{4p_{T}^{2}}$, $R_{i}(x,\mu_{i}^{2})$ is the nuclear modification function which reflect the nuclear shadowing effect~\cite{Eskola:2009uj},
$Z$ is the proton number, $N=A-Z$ is the neutron number and $A$ is the nucleon number.
$p_{i}(x,\mu^{2})$ and $n_{i}(x,\mu^{2})$ are the parton distributions of the protons and neutrons~\cite{Pumplin:2002vw}, respectively.

Denoting the virtual photo-absorption amplitude by $M^{\mu}$, the differential cross section in the parton level can be presented as
\begin{eqnarray}\label{sdpho.coh.dir}
&&d\sigma(\alpha+b\rightarrow \alpha+\gamma^{*}+d)\nonumber\\
&=&\!\!\!\frac{e_{\alpha}^{2}e^{2}}{Q^{2}}\frac{M^{\mu}M^{*\nu}\rho_{\mu\nu}}{4p_{\mathrm{CM}}\sqrt{s_{0}}}
\frac{d^{3}p_{\alpha'}}{(2\pi)^{3}2E_{\alpha'}}d\mathrm{PS}_{2}(q+p_{b};p_{c},p_{d}),
\end{eqnarray}
where $p_{\mathrm{CM}}$ and $\sqrt{s_{0}}$ are the momentum and energy of $\alpha b$ CM frame, respectively.
$e_{\alpha}$ is the charge of projectile $\alpha$.
$E_{\alpha'}$ is the energy of the scattered projectile, and we employ the short-hand notation
\begin{eqnarray}\label{dab.Gen1}
d\mathrm{PS}_{n}(P;p_{1},...,p_{n})=(2\pi)^{4}\delta^{4}(P-\sum_{i=1}^{n}p_{i})\prod_{i=1}^{n}\frac{d^{3}p_{i}}{(2\pi)^{3}2E_{i}},
\end{eqnarray}
for the Lorentz invariant N-particle phase-space element.
$\rho^{\mu\nu}$ is the density matrix of the virtual photon produced by projectile $\alpha$,
\begin{eqnarray}\label{Rou.}
\rho^{\mu\nu}&=&-\left(g^{\mu\nu}-\frac{q^{\mu}q^{\nu}}{q^{2}}\right)F_{2}(Q^{2})\nonumber\\
&&-\frac{(2P_{\alpha}-q)^{\mu}(2P_{\alpha}-q)^{\nu}}{q^{2}}F_{1}(Q^{2}),
\end{eqnarray}
$F_{1}(Q^{2})$ and $F_{2}(Q^{2})$ are the general notations for the form factors of projectile.

To obtain the $Q^{2}$ dependent cross section, it is convenient to do the calculations in the rest frame of $\alpha$, where $|\mathbf{q}|=|\mathbf{p}_{\alpha'}|=r$, $Q^{2}=-q^{2}=(p_{\alpha}-p_{\alpha'})^{2}=2m_{\alpha}(\sqrt{r^{2}+m_{\alpha}^{2}}-m_{\alpha})$,
$d^{3}p_{\alpha'}=r^{2}drd\cos\theta d\varphi$, and $y=(q\cdot p_{b})/(p_{\alpha}\cdot p_{b})=(q_{0}-|\mathbf{p}_{b}|r\cos\theta/E_{b})/m_{\alpha}$.
By using the Jacobian determinant,
\begin{eqnarray}\label{Jac. Q2y}
d\cos\theta dr=\left|\frac{D(r,\cos\theta)}{D(y,Q^{2})}\right|dydQ^{2}=\frac{E_{\alpha'}E_{b}}{2\left|\textbf{p}_{b}\right|r^{2}}dydQ^{2},
\end{eqnarray}
the cross section of subprocess $\alpha+b\rightarrow \alpha+\gamma^{*}+d$
can be further expressed as follow
\begin{eqnarray}\label{sdphodPS.coh.dir}
&&d\sigma(\alpha+b\rightarrow \alpha+\gamma^{*}+d)\nonumber\\
&=&\frac{e_{\alpha}^{2}e^{2}}{Q^2}
\frac{W^{\mu\nu}\rho_{\mu\nu}}{4(2\pi)^{3}\lambda'}\left(\frac{\lambda'}{\lambda}\frac{dydQ^{2}d\varphi}{\sqrt{1-f^{2}(s_{0},m_{\alpha},m_{b})}}\right),
\end{eqnarray}
with
\begin{eqnarray}\label{lambda}
&&\lambda=\sqrt{[s_{0}-(m_{\alpha}-m_{b})^{2}][s_{0}-(m_{\alpha}+m_{b})^2]},\nonumber\\
&&\lambda'=\sqrt{[\hat{s}-(m_{q}-m_{b})^{2}][\hat{s}-(m_{q}+m_{b})^2]},\nonumber\\
&&f(s_{0},m_{\alpha},m_{b})=\frac{2m_{\alpha}m_{b}}{s_{0}-m_{\alpha}^{2}-m_{b}^{2}},
\end{eqnarray}
where $W^{\mu\nu}=\frac{1}{2}\int M^{\mu}M^{*\nu}d\textrm{PS}_{2}(q+p_{b};p_{c},p_{b}')$.

In order to take into account explicitly gauge invariance, it is convenient to use the following linear combinations~\cite{Budnev:1974de}
\begin{eqnarray}\label{linear}
Q^{\mu}\!\!&=&\!\!\sqrt{\frac{-q^{2}}{(q\cdot p_{b})^{2}-q^{2}p_{b}^{2}}}\left(p_{b}^{\mu}-q^{\mu}\frac{q\cdot p_{b}}{q^{2}}\right),\nonumber\\
R^{\mu\nu}\!\!&=&\!\!-g^{\mu\nu}+\frac{(q\cdot p_{b})(q^{\mu}p_{b}^{\nu}+q^{\nu}p_{b}^{\mu})-q^{2}p_{b}^{\mu}p_{b}^{\nu}-p_{b}^{2}q^{\mu}q^{\nu}}{(q\cdot p_{b})^{2}-q^{2}p_{b}^{2}},\nonumber\\
\end{eqnarray}
they satisfy the relations: $q_{\mu}Q^{\mu}=q_{\mu}R^{\mu\nu}=0$, $Q^{\mu}Q_{\mu}=1$.
Having expended $W^{\mu\nu}$ in these tensors, we obtain
\begin{eqnarray}\label{W}
W^{\mu\nu}=R^{\mu\nu}W_{\mathrm{T}}+Q^{\mu}Q^{\nu}W_{\mathrm{S}}.
\end{eqnarray}
The dimensionless invariant functions $W_{\mathrm{T}}$ and $W_{\mathrm{S}}$ are simply connected with the cross section for transverse or scalar photon absorption $\sigma_{\mathrm{T}}$ and $\sigma_{\mathrm{S}}$ respectively:
\begin{eqnarray}\label{coh.dir.TL}
\sigma_{\mathrm{T}}(\gamma^{*}+b\rightarrow\gamma^{*}+d)&=&\frac{W_{\mathrm{T}}}{\lambda'},\nonumber\\
\sigma_{\mathrm{S}}(\gamma^{*}+b\rightarrow\gamma^{*}+d)&=&\frac{W_{\mathrm{S}}}{\lambda'}.
\end{eqnarray}
Thus the differential cross section of subprocess $\alpha+b\rightarrow \alpha+\gamma^{*}+d$ can finally be expressed as
\begin{eqnarray}\label{Gen.TL}
&&\!\!\frac{d\sigma}{dQ^{2}}(\alpha+b\rightarrow \alpha+\gamma^{*}+b)\nonumber\\
&=&\!\!dy\frac{e_{\alpha}^{2}\alpha_{\mathrm{em}}}{4\pi}\frac{y\rho_{\mu\nu}}{Q^{2}}[R^{\mu\nu}\sigma_{\mathrm{T}}(\gamma^{*}+b\rightarrow\gamma^{*}+d)+Q^{\mu}Q^{\nu}\nonumber\\
&\times&\!\! \sigma_{\mathrm{S}}(\gamma^{*}+b\rightarrow\gamma^{*}+d)]\left(\frac{\lambda'}{y\lambda}\frac{1}{\sqrt{1-f^{2}(s_{0},m_{\alpha},m_{b})}}\right)\nonumber\\
&=&\!\!dy\frac{e_{\alpha}^{2}\alpha_{\mathrm{em}}}{2\pi}\left[\frac{y\rho^{++}}{Q^{2}}\sigma_{\mathrm{T}}(\gamma^{*}+b\rightarrow\gamma^{*}+d)+\frac{y\rho^{00}}{Q^{2}}\right.\nonumber\\
&\times&\!\!\!\!\left.\frac{1}{2}\sigma_{\mathrm{S}}(\gamma^{*}+b\rightarrow\gamma^{*}+d)\right]\left(\frac{\lambda'}{y\lambda}\frac{1}{\sqrt{1-f^{2}(s_{0},m_{\alpha},m_{b})}}\right),\nonumber\\
\end{eqnarray}
where the electromagnetic coupling constant is chosen as $\alpha_{\mathrm{em}}=1/137$, and
\begin{align}\label{Rouzz00}
\rho^{++}\!\!&=\!\frac{R^{\mu\nu}\rho_{\mu\nu}}{2}=F_{1}(Q^{2})\left[\frac{2(1-y)}{y^{2}}-\frac{2m_{\alpha}^{2}}{Q^{2}}\right]+F_{2}(Q^{2}),\displaybreak[0]\nonumber\\
\rho^{00}\!\!&=\!Q^{\mu}Q^{\nu}\rho_{\mu\nu}=F_{1}(Q^{2})\frac{y^{2}+4(1-y)}{y^{2}}-F_{2}(Q^{2}).\displaybreak[0]
\end{align}

Finally, it is necessary to discuss the detailed expressions of the form factors for each photon emission mechanism.
For the case of coherent-photon emission, the projectile $\alpha$ is nucleus, and thus the general notations $F_{1}(Q^{2})$ and $F_{2}(Q^{2})$ in Eq. (\ref{Rouzz00}) turn into the elastic nucleus form factors accordingly.
If the projectile is proton: $m_{\alpha}=m_{p}$, $F_{1}(Q^{2})$ and $F_{2}(Q^{2})$ can be written as~\cite{Kniehl:1990iv}
\begin{align}\label{F12p}
&F_{1}^{\mathrm{coh}}(Q^{2})=\frac{G_{\mathrm{E}}^{2}(Q^{2})+(Q^{2}/4m_{p}^{2})G_{\mathrm{M}}^{2}(Q^{2})}{1+Q^{2}/4m_{p}^{2}},\displaybreak[0]\nonumber\\
&F_{2}^{\mathrm{coh}}(Q^{2})=G_{\mathrm{M}}^{2}(Q^{2}),\displaybreak[0]
\end{align}
where electric form factor $G_{\mathrm{E}}(Q^{2})$ can be parameterized by the dipole form: $G_{\mathrm{E}}(Q^{2})=1/(1+Q^{2}/0.71~\textrm{GeV})^{2}$, and the magnetic form factor is $G_{\mathrm{M}}(Q^{2})=2.793\ G_{\mathrm{E}}(Q^{2})$.
If the projectile is lead: $m_{\alpha}=m_{Pb}$, $F_{1}(Q^{2})$ and $F_{2}(Q^{2})$ are changed accordingly,
\begin{align}\label{F12Pb}
&F_{1}^{\mathrm{coh}}(Q^{2})=Z^{2}F_{\textrm{em}}^{2}(Q^2),\displaybreak[0]\nonumber\\
&F_{2}^{\mathrm{coh}}(Q^{2})=\mu^{2}F_{\textrm{em}}^{2}(Q^{2}),\displaybreak[0]
\end{align}
where
\begin{align}\label{FPbem}
F_{\textrm{em}}(Q^{2})&=\frac{3}{(QR_{A})^{3}}[\sin(QR_{A})\displaybreak[0]\nonumber\\
&-QR_{A}\cos(QR_{A})]\frac{1}{1+a^{2}Q^{2}},\displaybreak[0]
\end{align}
is the electromagnetic form factor parameterization from the STARlight MC generator~\cite{Dyndal:2019ylt}, 
in which $R_{A}=1.1A^{1/3}~\textrm{fm}$, $a=0.7~\textrm{fm}$ and $Q=\sqrt{Q^{2}}$.
It should be mentioned that in Martin-Ryskin method~\cite{Martin:2014nqa}, the square of the electric form factor is used as the coherent probability or weighting factor in p-p collision:  $w_{c}=G_{\mathrm{E}}^{2}(Q^2)$, while the effect of magnetic form factor is neglected.
In the present paper, we extend the central ideal of this method to deal with the photon emission processes in heavy-ion collisions, where the effect of magnetic form factor is also included.

For the case of incoherent emission, the projectile $\alpha$ is the parton inside the nucleus, and the 'remained' probability, $1-w_{c}$, has to be considered for avoiding double counting.
In p-p collision, the general notations $F_{1}(Q^{2})$ and $F_{2}(Q^{2})$ in Eq. (\ref{Rouzz00}) have the following forms
\begin{eqnarray}\label{L}
&&F_{1}^{\mathrm{incoh}}(Q^{2})=F_{2}^{\mathrm{incoh}}(Q^{2})=1-G_{\mathrm{E}}^{2}(Q^{2}).
\end{eqnarray}
While in p-Pb and Pb-Pb collisions, the incoherent reactions should further be distinguished as the ordinary-incoherent and ultra-incoherent photon emissions.
For ordinary-incoherent photon emission, the projectile is the proton inside the lead: $m_{\alpha}=m_{p}$, and $F_{1}(Q^{2})$ and $F_{2}(Q^{2})$ should be expressed as
\begin{eqnarray}\label{L.OIC}
&&F^{\textrm{OIC}}_{1}(Q^{2})=[1-F_{\mathrm{em}}^{2}(Q^2)]G_{\mathrm{E}}^{2}(Q^{2})\frac{4m_{p}^{2}+7.8Q^{2}}{4m_{p}^{2}+Q^{2}},\nonumber\\
&&F^{\textrm{OIC}}_{2}(Q^{2})=[1-F_{\mathrm{em}}^{2}(Q^2)]7.8G_{\mathrm{E}}^{2}(Q^{2}),
\end{eqnarray}
and for ultra-incoherent photon emission, the projectile is the quark inside the lead: $m_{\alpha}=m_{q}=0$, since the neutron can not emit photon coherently, the weighting factor for the proton
and neutron inside nucleus are different:
\begin{eqnarray}\label{L.UIC}
&&F^{\textrm{UIC}}_{1 p}(Q^{2})=F^{\textrm{UIC}}_{2 p}(Q^{2})=[1-F_{\mathrm{em}}^{2}(Q^2)][1-G_{\mathrm{E}}^{2}(Q^{2})],\nonumber\\
&&F^{\textrm{UIC}}_{1 n}(Q^{2})=F^{\textrm{UIC}}_{2 n}(Q^{2})=[1-F_{\mathrm{em}}^{2}(Q^2)].
\end{eqnarray}

\subsection{The $Q^{2}$ distribution of large $p_{T}$ dileptons production}
\label{subsec:Q2_dis.}

Since photons, dileptons and the dileptonic decay channel of light vector mesons do not participate in the strong interactions directly, their productions have long been proposed as ideal probes of quark-gluon plasma (QGP) properties.
In present section, we employ the accurate expression Eq. (\ref{Gen.TL}) to give the $Q^{2}$ dependent differential cross sections for large $p_{T}$ dileptons photoproduction.
Here large $p_{T}$ means that the transverse momentum of the final state is larger than $1~\mathrm{GeV}$.
In the initial state, the photoproduction processes may be direct and resolved~\cite{Ma:2018zzq}.
In the direct photoproduction processes, the high-energy photon, emitted from the projectile $\alpha$, interacts with the partons $b$ of target nucleus $B$ by the interactions of quark-photon Compton scattering.
In the resolved photoproduction processes, the uncertainty principle allows the high-energy hadron-like photon fluctuates into a color singlet state with multiple $q\bar{q}$ pairs and gluons.
Due to this fluctuation, the photon interacts with the partons in $B$ like a hadron, and the subprocesses are quark-antiquark annihilation and quark-gluon Compton scattering.
We must keep in mind that the distinction between these two types contributions does not really exist, only the sum of them has a physical meaning.
Actually, as always with photons, the situation is quite complex.
Together with three different photon emission mechanisms mentioned previously, we have six classes of processes: coherent-direct (coh.dir.), coherent-resolved (coh.res.), ordinary-incoherent direct (OIC.dir.), ordinary-incoherent resolved (OIC.res.), ultra-incoherent direct (UIC.dir.) and ultra-incoherent resolved (UIC.res.) processes.
These abbreviations will appear in many places of remained content and we do not explain its meaning again.

For the case of coherent-direct processes, the virtual photon emitted from the whole incident nucleus $A$ interacts
with parton $b$ of target nucleus $B$ via photon-quark Compton scattering, and
nucleus $A$ remains intact after photon emitted.
The differential cross section of large $p_{T}$ dileptons produced in this channel can be written as
\begin{eqnarray}\label{d.coh.dir.}
&&\frac{d\sigma^{\textrm{coh.dir.}}}{dQ^{2}}(A+B\rightarrow A+l^{+}l^{-}+X)\nonumber\\
&=&2\sum_{b}\int dM^{2}dx_{b}f_{b/B}(x_{b},\mu_{b}^{2})\frac{\alpha_{\mathrm{em}}}{3\pi M^{2}}\sqrt{1-\frac{4m_{l}^{2}}{M^{2}}}\nonumber\\
&\times&\left(1+\frac{2m_{l}^{2}}{M^{2}}\right)\frac{d\sigma}{dQ^{2}}(A+b\rightarrow A+\gamma^{*}+d),
\end{eqnarray}
where $M$ is the invariant mass of dileptons, $m_{l}$ is lepton mass.
The factor of two in Eq. (\ref{d.coh.dir.}) arises because both nuclei emit photons and thus serve as targets.
But for p-Pb collision the photon emitter can be either proton or lead, instead the factor of two,
these two contributions have to be summed together.

The partonic cross section $d\sigma(A+b\rightarrow A+\gamma^{*}+d)/dQ^{2}$ is the same as Eq. (\ref{Gen.TL}), and the expressions of the transverse and scalar photon  cross sections are
\begin{align}\label{dTL.dir.}
&\frac{d\hat{\sigma}_{\mathrm{T}}}{d\hat{t}}(\gamma^{*}+b\rightarrow \gamma^{*}+d)\displaybreak[0]\nonumber\\
&=\frac{2\pi\alpha_{\mathrm{em}}^{2}e_{b}^{4}}{(\hat{s}+Q^{2})^2}\left[-\frac{\hat{t}}{\hat{s}}-\frac{\hat{s}}{\hat{t}}-M^{2}Q^{2}\right.
\left(\frac{1}{\hat{s}^{2}}+\frac{1}{\hat{t}^{2}}\right)\displaybreak[0]\nonumber\\
&\left.+2(Q^{2}-M^{2})\frac{\hat{u}}{\hat{s}\hat{t}}\right]+4\pi\alpha_{\mathrm{em}}^{2}e_{b}^{4}
\frac{Q^{2}\hat{u}(\hat{t}-M^2)^2}{\hat{t}^{2}(\hat{s}+Q^{2})^4},\displaybreak[0]\nonumber\\
&\displaybreak[0]\nonumber\\
&\frac{d\hat{\sigma}_{\mathrm{S}}}{d\hat{t}}(\gamma^{*}+b\rightarrow \gamma^{*}+d)
=2\pi\alpha_{\mathrm{em}}^{2}e_{b}^{4}\frac{Q^{2}\hat{u}(\hat{t}-M^2)^2}{\hat{t}^{2}(\hat{s}+Q^{2})^4},\displaybreak[0]
\end{align}
where $e_{b}$ is the charge of massless quark $b$, $\hat{s}$, $\hat{t}$ and $\hat{u}$ are the Mandelstam variables and its detailed expressions for each case can be found in
Appendix \ref{FKR}.

For the case of ordinary-incoherent direct processes, the photon emitter is the proton $a$ inside the nucleus $A$, and the corresponding cross section is
\begin{eqnarray}\label{d.OIC.dir.}
&&\frac{d\sigma^{\textrm{OIC.dir.}}}{dQ^{2}}(A+B\rightarrow X_{A}+l^{+}l^{-}+X)\nonumber\\
&=&2Z_{Pb}\sum_{b}\int dM^{2}dx_{b}f_{b/B}(x_{b},\mu_{b}^{2})\frac{\alpha_{\mathrm{em}}}{3\pi M^{2}}\sqrt{1-\frac{4m_{l}^{2}}{M^{2}}}\nonumber\\
&\times&\left(1+\frac{2m_{l}^{2}}{M^{2}}\right)\frac{d\sigma}{dQ^{2}}(p+b\rightarrow p+\gamma^{*}+d).
\end{eqnarray}
And for the case of ultra-incoherent direct processes, the virtual photon emitted from the quarks $a$ inside nucleus $A$ interacts with parton $b$ of nucleus $B$ via the photon-quark interaction, and $A$ is allowed to break up after photon emitted.
Similarly, the corresponding differential cross section has the form of
\begin{eqnarray}\label{d.UIC.dir.}
&&\frac{d\sigma^{\textrm{UIC.dir.}}}{dQ^{2}}(A+B\rightarrow X_{A}+l^{+}l^{-}+X)\nonumber\\
&=&2\sum_{a,b}\int dM^{2}dx_{a}dx_{b}f_{a/A}(x_{a},\mu_{a}^{2})f_{b/B}(x_{b},\mu_{b}^{2})\frac{\alpha_{\mathrm{em}}}{3\pi M^{2}}\nonumber\\
&\times&\sqrt{1-\frac{4m_{l}^{2}}{M^{2}}}\left(1+\frac{2m_{l}^{2}}{M^{2}}\right)\frac{d\sigma}{dQ^{2}}(a+b\rightarrow a+\gamma^{*}+d),\nonumber\\
\end{eqnarray}
where $x_{a}=p_{a}/P_{A}$ is parton's momentum fraction, $f_{a/A}(x_{a},\mu_{a}^{2})$ is the parton distribution function of nucleus $A$, $\mu_{a}=\sqrt{4p_{T}^{2}}$, and the cross section of the partonic processes $a + b\rightarrow a + \gamma^{*} +d$ can be derived from Eqs. (\ref{Gen.TL}) and (\ref{dTL.dir.}) with $m_{\alpha}=m_{q}=0$ and $e_{\alpha}=e_{a}$, where $e_{a}$ is the charge of massless quark $a$.

In the coherent-resolved processes, the incident nucleus $A$ emits a high energy virtual photon, then the parton $a'$ from the resolved photon interacts with the parton $b$ from another incident nucleus $B$ via quark-antiquark annihilation or quark-gluon Compton scattering, and $A$ remains intact after photon emitted.
The relevant differential cross section is:
\begin{eqnarray}\label{coh.res.}
&&\frac{d\sigma^{\mathrm{coh.res.}}}{dQ^{2}}(A+B\rightarrow A+l^{+}l^{-}+X)\nonumber\\
&=&\!\!2\sum_{b}\sum_{a'}\int dM^{2}dydx_{b}dz_{a'}d\hat{t}f_{b/B}(x_{b},\mu_{b}^{2})f_{\gamma}(z_{a'},\mu_{\gamma}^{2})\nonumber\\
&\times&\!\!\!\!\frac{e_{\alpha}^{2}\alpha_{\mathrm{em}}}{2\pi}\frac{y\rho^{++}_{\mathrm{coh}}}{Q^{2}}\frac{\alpha_{\mathrm{em}}}{3\pi M^{2}}\sqrt{1-\frac{4m_{l}^{2}}{M^{2}}}\left(1+\frac{2m_{l}^{2}}{M^{2}}\right)\frac{d\sigma_{a'b\rightarrow \gamma^{*}d}}{d\hat{t}},\nonumber\\
\end{eqnarray}
where $f_{\gamma}(z_{a'},\mu_{\gamma}^{2})$ is the parton distribution function of the resolved photon \cite{Gluck:1999ub}, $\mu_{\gamma}=\sqrt{4p_{T}^{2}}$, $z_{a'}$ denotes the parton's momentum fraction of the resolved photon emitted from the nucleus $A$.
The involved subprocesses are $q_{a'}\bar{q}_{b}\rightarrow\gamma^{*}g$, $q_{a'}g_{b}\rightarrow\gamma^{*}q$ and $g_{a'}q_{b}\rightarrow\gamma^{*}q$, its cross sections  can be found in Ref. \cite{Owens:1986mp}.

In the ordinary-incoherent resolved processes, the emitter of resolved virtual photon is the protons inside nucleus, and the corresponding cross section is
\begin{eqnarray}\label{OIC.res.pPb}
&&\frac{d\sigma^{\mathrm{OIC.res.}}}{dQ^{2}}(A+B\rightarrow X_{A}+l^{+}l^{-}+X)\nonumber\\
&=&\!\!2Z_{Pb}\sum_{b}\sum_{a'}\int dM^{2}dydx_{b}dz_{a'}d\hat{t}f_{b/B}(x_{b},\mu_{b}^{2})\nonumber\\
&\times&\!\! f_{\gamma}(z_{a'},\mu_{\gamma}^{2})\frac{\alpha_{\mathrm{em}}}{2\pi}\frac{y\rho^{++}_{\mathrm{OIC}}}{Q^{2}}\frac{\alpha_{\mathrm{em}}}{3\pi M^{2}}\sqrt{1-\frac{4m_{l}^{2}}{M^{2}}}\left(1+\frac{2m_{l}^{2}}{M^{2}}\right)\nonumber\\
&\times&\frac{d\sigma_{a'b\rightarrow \gamma^{*}d}}{d\hat{t}}.
\end{eqnarray}
And in the ultra-incoherent resolved processes, the quarks inside nucleus $A$ emit a hadron-like virtual photon, then the parton $a^\prime$ of this resolved photon interacts with parton $b$ inside nucleus $B$, and $A$ is break up after photon emitted.
The relevant differential cross section is
\begin{eqnarray}\label{incoh.res.}
&&\frac{d\sigma^{\mathrm{UIC.res.}}}{dQ^{2}}(A+B\rightarrow X_{A}+l^{+}l^{-}+X)\nonumber\\
&=&2\sum_{a,b}\sum_{a'}\int dM^{2}dydx_{a}dx_{b}dz_{a'}d\hat{t}f_{a/A}(x_{a},\mu_{a}^{2})\nonumber\\
&\times&\!\! f_{b/B}(x_{b},\mu_{b}^{2})f_{\gamma}(z_{a'},\mu_{\gamma}^{2})e_{a}^{2}\frac{\alpha_{\mathrm{em}}}{2\pi}\frac{y\rho^{++}_{\textrm{UIC}}}{Q^{2}}\frac{\alpha_{\mathrm{em}}}{3\pi M^{2}}\sqrt{1-\frac{4m_{l}^{2}}{M^{2}}}\nonumber\\
&\times&\!\! \left(1+\frac{2m_{l}^{2}}{M^{2}}\right)\frac{d\sigma_{a'b\rightarrow \gamma^{*}d}}{d\hat{t}}.
\end{eqnarray}

\subsection{The $p_{T}$ and $y_{r}$ distributions of large $p_{T}$ dileptons production}
\label{subsec:Dileptons_PT_yr}

The distributions in $p_{T}$ and the rapidity $y_{r}$ can be obtained by using the Jacobian determinant.
It needs to be emphasized that one should add a term with the exchange $(y_r\rightarrow -y_r)$ in the formulae of $y_{r}$ distribution, which reflects the fact that each colliding nucleus can serve as a photon emitter and as a target.

In the final state, the photoproduction of large $p_{T}$ dileptons can be divided into two categories: direct dileptons produced from a direct final photon which directly coupled to a quark of the hard subprocess,
fragmentation dileptons produced by the bremsstrahlung emitted from the final state partons [1].
In the following we will take into account all these aspects.

\subsubsection{Large $p_{T}$ direct dileptons production}
\label{sec:D_Dilepton_PT_yr}

It is straightforward to obtain the distributions in $p_{T}$ and $y_{r}$, by accordingly reordering and redefining the involved integration variables.
At the beginning, the Mandelstam variables should be written in the forms
\begin{align}\label{Mant.yr}
\hat{s}&=m_{b}^{2}-M^{2}+2\cosh y_{r}M_{T}\sqrt{\hat{s}},\nonumber\\
\hat{t}&=M^{2}-Q^{2}-2M_{T}\left[\hat{E}_{\gamma}\cosh y_{r}-\hat{p}_{\mathrm{CM}}\sinh y_{r}\right],\nonumber\\
\hat{u}&=M^{2}+m_{b}^{2}-2M_{T}\left[\hat{E}_{b}\cosh y_{r}+\hat{p}_{\mathrm{CM}}\sinh y_{r}\right],
\end{align}
where $y_{r}=(1/2)\ln(E+p_{z})/(E-p_{z})$, $M_{T}=\sqrt{p_{T}^{2}+M^{2}}$ is the dilepton transverse mass.
$\hat{E}_{\gamma}=(\hat{s}-Q^{2}-m_{b}^{2})/(2\sqrt{\hat{s}})$, $\hat{E}_{b}=(\hat{s}+m_{b}^{2}+Q^{2})/(2\sqrt{\hat{s}})$ and $\hat{p}_{\mathrm{CM}}=\sqrt{[(\hat{s}+Q^{2}-m_{b}^{2})^{2}+4Q^{2}m_{b}^{2}]}/(2\sqrt{\hat{s}})$ are the energies and momentum in $\gamma^{*}b$ CM frame.

For the case of direct photoproduction processes, the variables $x_{b}$ and $\hat{t}$ should be transformed into the following form by using the Jacobian determinant,
\begin{eqnarray}\label{Jac.dir.}
d\hat{t}dx_{b}=\mathcal{J}dy_{r}dp_{T}=\left|\frac{D(x_{b},\hat{t})}{D(y_{r},p_{T})}\right|dy_{r}dp_{T}.
\end{eqnarray}
Thus the corresponding differential cross sections of large $p_{T}$ direct dileptons production can be expressed as
\begin{align}
&\frac{d\sigma^{\textrm{coh.dir.}}}{dy_{r}dp_{T}}(A+B\rightarrow A+l^{+}l^{-}+X)\displaybreak[0]\nonumber\\
=&2\sum_{b}\int dM^{2}dQ^{2}dyf_{b/B}(x_{b},\mu_{b}^{2})\mathcal{J}\frac{\alpha_{\mathrm{em}}}{3\pi M^{2}}\sqrt{1-\frac{4m_{l}^{2}}{M^{2}}}\displaybreak[0]\nonumber\\
\times&\left(1+\frac{2m_{l}^{2}}{M^{2}}\right)\frac{d\sigma}{dQ^{2}dyd\hat{t}}(A+b\rightarrow A+\gamma^{*}+d),\label{ddPT.coh.dir}\displaybreak[0]\\
&\displaybreak[0]\nonumber\\
&\frac{d\sigma^{\textrm{OIC.dir.}}}{dy_{r}dp_{T}}(A+B\rightarrow X_{A}+l^{+}l^{-}+X)\displaybreak[0]\nonumber\\
=&2Z_{Pb}\sum_{b}\int dM^{2}dQ^{2}dyf_{b/B}(x_{b},\mu_{b}^{2})\mathcal{J}\frac{\alpha_{\mathrm{em}}}{3\pi M^{2}}\sqrt{1-\frac{4m_{l}^{2}}{M^{2}}}\displaybreak[0]\nonumber\\
\times&\left(1+\frac{2m_{l}^{2}}{M^{2}}\right)\frac{d\sigma}{dQ^{2}dyd\hat{t}}(p+b\rightarrow p+\gamma^{*}+d),\label{ddPTpPb.OIC.dir}\displaybreak[0]\\
&\displaybreak[0]\nonumber\\
&\frac{d\sigma^{\textrm{UIC.dir.}}}{dy_{r}dp_{T}}(A+B\rightarrow X_{A}+l^{+}l^{-}+X)\displaybreak[0]\nonumber\\
=&2\sum_{a,b}\int dM^{2}dQ^{2}dydx_{a}f_{a/A}(x_{a},\mu_{a}^{2})f_{b/B}(x_{b},\mu_{b}^{2})\mathcal{J}\frac{\alpha_{\mathrm{em}}}{3\pi M^{2}}\displaybreak[0]\nonumber\\
\times&\sqrt{1-\frac{4m_{l}^{2}}{M^{2}}}\left(1+\frac{2m_{l}^{2}}{M^{2}}\right)\frac{d\sigma}{dQ^{2}dyd\hat{t}}(a+b\rightarrow a+\gamma^{*}+d),
\label{ddPT.incoh.dir}\displaybreak[0]
\end{align}

For the case of resolved contributions, we should choose the variables $\hat{t}_{\gamma}$ and $z_{a'}$ to do the similar transformation,
\begin{eqnarray}\label{Jac.res.}
d\hat{t}_{\gamma}dz_{a'}=\mathcal{J}dy_{r}dp_{T}=\left|\frac{D(z_{a'},\hat{t}_{\gamma})}{D(y_{r},p_{T})}\right|dy_{r}dp_{T},
\end{eqnarray}
the corresponding differential cross sections are
\begin{align}
&\frac{d\sigma^{\mathrm{coh.res.}}}{dy_{r}dp_{T}}(A+B\rightarrow A+l^{+}l^{-}+X)\displaybreak[0]\nonumber\\
=&2\sum_{b}\sum_{a'}\int dM^{2}dQ^{2}dydx_{b}f_{b/B}(x_{b},\mu_{b}^{2})f_{\gamma}(z_{a'},\mu_{\gamma}^{2})\mathcal{J}\displaybreak[0]\nonumber\\
\times& e_{\alpha}^{2}\frac{\alpha_{\mathrm{em}}}{2\pi}\frac{y\rho^{++}_{\mathrm{coh}}}{Q^{2}}
\frac{\alpha_{\mathrm{em}}}{3\pi M^{2}}\sqrt{1-\frac{4m_{l}^{2}}{M^{2}}}\left(1+\frac{2m_{l}^{2}}{M^{2}}\right)\frac{d\sigma_{a'b\rightarrow \gamma^{*}d}}{d\hat{t}},\label{ddPT.coh.res.}\displaybreak[0]\\
&\displaybreak[0]\nonumber\\
&\frac{d\sigma^{\mathrm{OIC.res.}}}{dy_{r}dp_{T}}(A+B\rightarrow X_{A}+l^{+}l^{-}+X)\displaybreak[0]\nonumber\\
=&2Z_{Pb}\sum_{b}\sum_{a'}\int dM^{2}dQ^{2}dydx_{b}f_{b/B}(x_{b},\mu_{b}^{2})f_{\gamma}(z_{a'},\mu_{\gamma}^{2})\displaybreak[0]\nonumber\\
\times&\mathcal{J}\frac{\alpha_{\mathrm{em}}}{2\pi}\frac{y\rho^{++}_{\mathrm{OIC}}}{Q^{2}}\frac{\alpha_{\mathrm{em}}}{3\pi M^{2}}\sqrt{1-\frac{4m_{l}^{2}}{M^{2}}}\left(1+\frac{2m_{l}^{2}}{M^{2}}\right)\frac{d\sigma_{a'b\rightarrow \gamma^{*}d}}{d\hat{t}},\label{ddPTpPb.OIC.res.}\displaybreak[0]\\
&\displaybreak[0]\nonumber\\
&\frac{d\sigma^{\mathrm{UIC.res.}}}{dy_{r}dp_{T}}(A+B\rightarrow X_{A}+l^{+}l^{-}+X)\displaybreak[0]\nonumber\\
=&2\sum_{a,b}\sum_{a'}\int dM^{2}dQ^{2}dydx_{a}dx_{b}f_{a/A}(x_{a},\mu_{a}^{2})f_{b/B}(x_{b},\mu_{b}^{2})\displaybreak[0]\nonumber\\
\times& f_{\gamma}(z_{a'},\mu_{\gamma}^{2})\mathcal{J}e_{a}^{2}\frac{\alpha_{\mathrm{em}}}{2\pi}\frac{y\rho^{++}_{\mathrm{UIC}}}{Q^{2}}\frac{\alpha_{\mathrm{em}}}{3\pi M^{2}}\sqrt{1-\frac{4m_{l}^{2}}{M^{2}}}\left(1+\frac{2m_{l}^{2}}{M^{2}}\right)\displaybreak[0]\nonumber\\
\times&\frac{d\sigma_{a'b\rightarrow \gamma^{*}d}}{d\hat{t}},\label{ddPT.incoh.res.}\displaybreak[0]
\end{align}
where the Mandelstam variables of resolved photoproduction processes are the same as Eq. (\ref{Mant.yr}) but for $Q^{2}=0$.

\subsubsection{Large $p_{T}$ fragmentation dileptons production}
\label{sec:F_Dilepton_PT_yr}

The fragmentation dileptons production is also an important channel which involves a perturbative part - the bremsstrahlung of the virtual photon- and a nonperturbative part,  which is described by the dilepton fragmentation function~\cite{Kang:2008wv}
\begin{align}\label{f.frag}
&D_{q_{c}}^{l^{+}l^{-}}(z_{c},M^{2},Q^{2})\displaybreak[0]\nonumber\\
=&\frac{\alpha_{\mathrm{em}}}{3\pi M^{2}}\sqrt{1-\frac{4m_{l}^{2}}{M^{2}}}(1+\frac{2m_{l}^{2}}{M^{2}})D_{q_{c}}^{\gamma^{*}}(z_{c},Q^{2}),\displaybreak[0]
\end{align}
where $z_{c}=2p_{T}\cosh y_{r}/\sqrt{\hat{s}}$ is the momentum fraction of the final state dileptons, $D_{q_{c}}^{\gamma^{*}}(z_{c},Q^{2})$ is the virtual photon fragmentation function.

First of all, we should rewrite the Mandelstam variables as following forms for fragmentation dileptons production,
\begin{eqnarray}\label{Mant.coh.dir.frag.}
&&\hat{s}=\frac{yx_{b}s}{N_{A}}-Q^{2},\nonumber\\
&&\hat{t}=-Q^{2}-\frac{\hat{s}}{2\cosh(y_{r})}e^{-y_{r}}+\frac{Q^{2}}{2\cosh(y_{r})}e^{y_{r}},\nonumber\\
&&\hat{u}=-\frac{\hat{s}}{2\cosh(y_{r})}e^{y_{r}}-\frac{Q^{2}}{2\cosh(y_{r})}e^{y_{r}}.
\end{eqnarray}
In this case, the variables $z_{c}$ and $\hat{t}$ should be chosen to do the transformation
\begin{eqnarray}\label{Jac.frag}
d\hat{t}dz_{c}=\mathcal{J}dy_{r}dp_{T}=\left|\frac{D(z_{c},\hat{t})}{D(y_{r},p_{T})}\right|dy_{r}dp_{T}.
\end{eqnarray}

For the case of coherent-direct processes, the corresponding differential cross section can be expressed as
\begin{eqnarray}\label{ddPT.coh.dir.frag}
&&\frac{d\sigma^{\textrm{coh.dir.-frag.}}}{dy_{r}dp_{T}}(A+B\rightarrow A+l^{+}l^{-}+X)\nonumber\\
&=&2\sum_{b,c}\int dM^{2}dQ^{2}dydx_{b}f_{b/B}(x_{b},\mu_{b}^{2})D_{q_{c}}^{l^{+}l^{-}}(z_{c},Q^{2})\nonumber\\
&&\times \frac{\mathcal{J}}{z_{c}}\frac{d\sigma}{dQ^{2}dyd\hat{t}}(A+b\rightarrow A+c+d),
\end{eqnarray}
where the cross section $d\sigma(A+b\rightarrow A+c+d)/dQ^{2}$ has been discussed in Eq. (\ref{Gen.TL}), and the partonic subprocesses involved in this channel are $q\gamma^{*}\rightarrow q\gamma$, $q\gamma^{*}\rightarrow qg$ and $g\gamma^{*}\rightarrow q\bar{q}$.
For $q\gamma^{*}\rightarrow q\gamma$, its cross sections are the same as Eq.~(\ref{dTL.dir.}), but for $M^{2}=0$.
While for $q\gamma^{*}\rightarrow qg$, the transverse and scalar photon cross sections are calculated in the following
\begin{align}\label{dTL.dir.qg}
&\frac{d\hat{\sigma}_{\mathrm{T}}}{d\hat{t}}(\gamma^{*}+q\rightarrow g+q)\displaybreak[0]\nonumber\\
=&\frac{8\pi\alpha_{\mathrm{em}}\alpha_{s}e_{q}^{2}}{3(\hat{s}+Q^{2})^2}\left[-\frac{\hat{t}}{\hat{s}}-\frac{\hat{s}}{\hat{t}}+2Q^{2}
\frac{\hat{u}}{\hat{s}\hat{t}}\right]+\frac{16\pi\alpha_{\mathrm{em}}\alpha_{s}e_{q}^{2}}{3}\frac{Q^{2}\hat{u}}{(\hat{s}+Q^{2})^4},\displaybreak[0]\nonumber\\
&\displaybreak[0]\nonumber\\
&\frac{d\hat{\sigma}_{\mathrm{S}}}{d\hat{t}}(\gamma^{*}+q\rightarrow g+q)=\frac{8\pi\alpha_{\mathrm{em}}\alpha_{s}e_{q}^{2}}{3}\frac{Q^{2}\hat{u}}{(\hat{s}+Q^{2})^4},
\displaybreak[0]
\end{align}
and also for the subprocess $g\gamma^{*}\rightarrow q\bar{q}$,
\begin{align}\label{dTL.dir.qqba}
&\frac{d\hat{\sigma}_{\mathrm{T}}}{d\hat{t}}(\gamma^{*}+g\rightarrow q+\bar{q})\displaybreak[0]\nonumber\\
=&\frac{\pi\alpha_{\mathrm{em}}\alpha_{s}e_{q}^{2}}{(\hat{s}+Q^{2})^2}\left[\frac{\hat{t}}{\hat{u}}+\frac{\hat{u}}{\hat{t}}-2Q^{2}
\frac{\hat{s}}{\hat{u}\hat{t}}\right]-\frac{2\pi\alpha_{\mathrm{em}}\alpha_{s}e_{q}^{2}}{(\hat{s}+Q^{2})^2}\frac{Q^{2}\hat{s}}{(\hat{u}+Q^{2})^2},\displaybreak[0]\nonumber\\
&\displaybreak[0]\nonumber\\
&\frac{d\hat{\sigma}_{\mathrm{S}}}{d\hat{t}}(\gamma^{*}+g\rightarrow q+\bar{q})
=-\frac{\pi\alpha_{\mathrm{em}}\alpha_{s}e_{q}^{2}}{(\hat{s}+Q^{2})^2}\frac{Q^{2}\hat{s}}{(\hat{u}+Q^{2})^2}.\displaybreak[0]
\end{align}

For the cases of ordinary- and ultra-incoherent direct processes, the differential cross sections are
\begin{align}
&\frac{d\sigma^{\textrm{OIC.dir.-frag.}}}{dy_{r}dp_{T}}(A+B\rightarrow X_{A}+l^{+}l^{-}+X)\displaybreak[0]\nonumber\\
=&2Z_{Pb}\sum_{b,c}\int dM^{2}dQ^{2}dydx_{b}f_{b/B}(x_{b},\mu_{b}^{2})D_{q_{c}}^{l^{+}l^{-}}(z_{c},Q^{2})\displaybreak[0]\nonumber\\
\times& \frac{\mathcal{J}}{z_{c}}\frac{d\sigma}{dQ^{2}dyd\hat{t}}(p+b\rightarrow p+c+d),\label{ddPTpPb.OIC.dir.frag}\displaybreak[0]\\
&\displaybreak[0]\nonumber\\
&\frac{d\sigma^{\textrm{UIC.dir.-frag.}}}{dy_{r}dp_{T}}(A+B\rightarrow X_{A}+l^{+}l^{-}+X)\displaybreak[0]\nonumber\\
=&2\sum_{a,b,c}\int dM^{2}dQ^{2}dydx_{a}dx_{b}f_{a/A}(x_{a},\mu_{a}^{2})f_{b/B}(x_{b},\mu_{b}^{2})\displaybreak[0]\nonumber\\
\times& D_{q_{c}}^{l^{+}l^{-}}(z_{c},Q^{2})\frac{\mathcal{J}}{z_{c}}\frac{d\sigma}{dQ^{2}dyd\hat{t}}(a+b\rightarrow c+d).
\label{ddPT.incoh.dir.frag}\displaybreak[0]
\end{align}

For the case of resolved contributions, the differential cross sections of large $p_{T}$ fragmentation dileptons can be presented as
\begin{align}
&\frac{d\sigma^{\mathrm{coh.res.-frag.}}}{dy_{r}dp_{T}}(A+B\rightarrow A+l^{+}l^{-}+X)\displaybreak[0]\nonumber\\
=&2\sum_{b}\sum_{a',c}\int dM^{2}dQ^{2}dydx_{b}dz_{a'}f_{b/B}(x_{b},\mu_{b}^{2})f_{\gamma}(z_{a'},\mu_{\gamma}^{2})\displaybreak[0]\nonumber\\
\times&D_{q_{c}}^{l^{+}l^{-}}(z_{c},Q^{2})\frac{\mathcal{J}}{z_{c}}e_{\alpha}^{2}\frac{\alpha_{\mathrm{em}}}{2\pi}\frac{y\rho^{++}_{\mathrm{coh}}}{Q^{2}}
\frac{d\sigma_{a'b\rightarrow cd}}{d\hat{t}},\label{ddPT.coh.res.frag}\displaybreak[0]\\
\displaybreak[0]\nonumber\\
&\frac{d\sigma^{\mathrm{OIC.res.-frag.}}}{dy_{r}dp_{T}}(A+B\rightarrow X_{A}+l^{+}l^{-}+X)\displaybreak[0]\nonumber\\
=&2Z_{Pb}\sum_{b}\sum_{a',c}\int dM^{2}dQ^{2}dydx_{b}dz_{a'}f_{b/B}(x_{b},\mu_{b}^{2})\displaybreak[0]\nonumber\\
\times&f_{\gamma}(z_{a'},\mu_{\gamma}^{2})D_{q_{c}}^{l^{+}l^{-}}(z_{c},Q^{2})\frac{\mathcal{J}}{z_{c}}\frac{\alpha_{\mathrm{em}}}{2\pi}
\frac{y\rho^{++}_{\mathrm{OIC}}}{Q^{2}}\frac{d\sigma_{a'b\rightarrow cd}}{d\hat{t}},\label{ddPTpPb.OIC.res.frag}\displaybreak[0]\\
\displaybreak[0]\nonumber\\
&\frac{d\sigma^{\mathrm{UIC.res.-frag.}}}{dy_{r}dp_{T}}(A+B\rightarrow X_{A}+l^{+}l^{-}+X)\displaybreak[0]\nonumber\\
=&2\sum_{a,b}\sum_{a',c}\int dM^{2}dQ^{2}dydx_{a}dx_{b}dz_{a'}f_{a/A}(x_{a},\mu_{a}^{2})\displaybreak[0]\nonumber\\
\times& f_{b/B}(x_{b},\mu_{b}^{2})f_{\gamma}(z_{a'},\mu_{\gamma}^{2})D_{q_{c}}^{l^{+}l^{-}}(z_{c},Q^{2})\frac{\mathcal{J}}{z_{c}}e_{a}^{2}
\frac{\alpha_{\mathrm{em}}}{2\pi}\frac{y\rho^{++}_{\mathrm{UIC}}}{Q^{2}}\displaybreak[0]\nonumber\\
\times& \frac{d\sigma_{a'b\rightarrow cd}}{d\hat{t}},
\displaybreak[0]\label{ddPT.incoh.res.frag}\displaybreak[0]
\end{align}
where the involved subprocesses are $qq\rightarrow qq$, $qq'\rightarrow qq'$,
$q\bar{q}\rightarrow q\bar{q}$, $q\bar{q}\rightarrow q'\bar{q}'$, $q\bar{q}'\rightarrow q\bar{q}'$,
$qg\rightarrow q\gamma$, $qg\rightarrow qg$ and $gg\rightarrow q\bar{q}$ \cite{Owens:1986mp}.
The Mandelstam variables of resolved contributions are the same as Eq. (\ref{Mant.coh.dir.frag.}) but for $Q^{2}=0$.

\subsection{Photoproduction of large $p_{T}$ photons and light vector mesons}
\label{photons LVDs}

In relativistic nucleus-nucleus collisions, a complicated hadronic system with a large multiplicity of particles is formed, involving the possibility to forme a phase of QCD matter -QGP which exists at extremely high temperature and density.
The ALICE experiment has been designed to study the physics of this QCD phase via heavy-ion collisions.
The photons and light mesons $(\rho, \omega,$ and $\phi)$ appear to be sensitive probes of QGP, which can be used to extract the key information on this matter.
Photons couple weakly to charged particles and not at all to themselves, so they are ideal tools for precision measurements.
They do not participate in the strong interaction directly, thus the photons do not likely suffer further collisions after they are produced.
And for light vector mesons, the strangeness enhancement can be accessed through the measurement of $\phi$ meson production, while the measurement of the $\rho$ spectral function can be used to reveal in-medium modifications of hadron properties close to the QCD phase boundary.
Moreover, it is interesting by itself, since it provides insight into soft QCD processes in the LHC energy range~\cite{ALICE:2011ad}.
Calculations in this regime are based on QCD inspired phenomenological models that must be tuned to data~\cite{Incani:2012ng}.

The photons and light vector mesons productions have received many studies within EPA~\cite{Klein:2017nqo}.
In this section, we would like to extend the photoproduction mechanism to study the production of large $p_{T}$ photons and the electromagnetic fragmentation production of the light vector mesons in p-p, p-Pb and Pb-Pb collisions.
The invariant cross sections of real photons production can be directly derived from those of dileptons production if the invariant mass of dileptons is zero $(M^{2}=0)$.
And for the electromagnetic fragmentation production of the light vector mesons, we adopt the following electromagnetic fragmentation function $D_{\gamma\rightarrow V}$ for a photon splitting to a light vector meson~\cite{Fleming:1994iu}:
\begin{eqnarray}\label{F.Efrag}
D_{\gamma\rightarrow V}=\frac{3\Gamma_{V\rightarrow e^{+}e^{-}}}{\alpha_{\mathrm{em}} m_{v}}
\end{eqnarray}
where $m_{v}$ is the vector meson's mass, $\Gamma_{V\rightarrow e^{+}e^{-}}$ is the electronic width.

\section{Equivalent photon spectrum}
\label{EPASp}

The idea of EPA was first developed by Fermi~\cite{Fermi:1924tc},
and was extended to include the interaction of relativistic charged particles by Weizs\"{a}cker and Williams, and the method now known as the Weizs\"{a}cker-Williams method (WWM) \cite{vonWeizsacker:1934nji}.
EPA as a useful technique, has been widely applied to obtain various cross sections for charged particles production in relativistic heavy-ion collisions~\cite{Budnev:1974de}.
And its application range has been extended beyond the realm of QED, such as equivalent pion method which describes the subthreshold pion production in nucleus-nucleus collision~\cite{Pirner:1980rn};
the nuclear WWM which describes excitation processes induced by the nuclear interaction in peripheral collisions of heavy ions~\cite{Feshbach:1976uu};
and a non-Abelian WWM describing the boosted gluon distribution functions in nucleus-nucleus collision~\cite{McLerran:1994vd}.
Although tremendous successes have been achieved, the discussion about the accuracy of EPA and its applicability range are still insufficient.
A number of imprecise statements pertaining to the essence and the advantages of EPA were given~\cite{Zhu:2015via, Zhu:2015qoz, Fu:2011zzm, Fu:2011zzf, Chin.Phys.C_36_721, Yu:2015kva, Yu:2017rfi, Yu:2017pot}.
For example, some improper kinematical bounds are widely used in the calculations~\cite{Kniehl:2001tk, Kniehl:1990iv, Drees:1988pp, Drees:1994zx, Drees:1989vq, sp};
the integration of some widely adopted spectra are performed over the entire kinematically allowed region, which leads to erroneous expressions;
EPA is applied to the processes which are essentially inapplicable~\cite{Drees:1994zx};
The serious double counting exists when the different photon emission mechanisms are considered simultaneously~\cite{Zhu:2015qoz, Fu:2011zzm, Fu:2011zzf, Chin.Phys.C_36_721, Yu:2015kva, Yu:2017rfi, Yu:2017pot}.

We have developed the exact treatment for photoproduction processes in heavy-ion collisions in section~\ref{sec:GeneralF_ET}, which can reduce to EPA by taking $Q^{2}\rightarrow 0$.
Detailed discussion can be also found in Ref.~\cite{Budnev:1974de}.
In present section we switch the accurate expression Eq. (\ref{Gen.TL}) to EPA form, which provides us a powerful and overall approach to study the features of EPA in heavy-ion collisions. 
In addition, a number of widely employed photon spectra are discussed.
The EPA consists in ignoring the fact that the photon in the photo-absorption amplitude is off mass shell and no longer transversely polarized from real photo-absorption.
As a result, the photoproduction processes can be factorized in terms of the real photo-absorption cross section and the equivalent photon spectrum.
Therefore, when switching to the approximate formulae of EPA, two simplifications should be performed.
Firstly, the scalar photon contribution $\sigma_{\mathrm{S}}$ is neglected;
secondly, the term of $\sigma_{\mathrm{T}}$ is substituted by its on-shell value.

Taking $Q^{2}\rightarrow0$, the linear combinations in Eq.~(\ref{linear}) can reduce to
\begin{align}\label{epsi.TL}
&\lim_{Q^{2}\rightarrow0}Q^{\mu}Q^{\nu}=-\varepsilon^{\mu\nu}_{\mathrm{S}}=-\frac{q^{\mu}q^{\nu}}{q^{2}},\displaybreak[0]\nonumber\\
&\lim_{Q^{2}\rightarrow0}R^{\mu\nu}=\varepsilon^{\mu\nu}_{\mathrm{T}}=-g^{\mu\nu}+\frac{(q^{\mu}p_{b}^{\nu}+q^{\nu}p_{b}^{\mu})}{q\cdot p_{b}},\displaybreak[0]
\end{align}
since gauge invariant $q^{\mu}W_{\mu\nu}=0$, the EPA form of the cross section in Eq. (\ref{Gen.TL}) can be written as:
\begin{align}\label{Gen.EPA}
&\!\!\lim_{Q^{2}\rightarrow0}\frac{d\sigma}{dy}(\alpha+b\rightarrow \alpha+\gamma^{*}+d)\displaybreak[0]\nonumber\\
=&\left(dQ^{2}\frac{e_{\alpha}^{2}\alpha_{\mathrm{em}}}{2\pi}\frac{y\rho^{++}}{Q^{2}}\right)\sigma_{\mathrm{T}}(\gamma^{*}+b\rightarrow\gamma^{*}+d)\bigg|_{Q^{2}=0}
\displaybreak[0]\nonumber\\
=&df_{\gamma}(y)\sigma_{\mathrm{T}}(\gamma^{*}+b\rightarrow\gamma^{*}+d)\bigg|_{Q^{2}=0},\displaybreak[0]
\end{align}
$f_{\gamma}(y)$ is the most general form of equivalent photon spectrum which is associated with various particles,
\begin{align}\label{fgamma.Gen.}
&\frac{df_{\gamma}(y)}{dQ^{2}}\displaybreak[0]\nonumber\\
=&\frac{\alpha_{\mathrm{em}}}{2\pi}\frac{y}{Q^{2}}\Bigg\{F_{1}(Q^{2})\left[\frac{2(1-y)}{y^{2}}-\frac{2m_{\alpha}^{2}}{Q^{2}}\right]+F_{2}(Q^{2})\Bigg\}\displaybreak[0]\nonumber\\
\approx&\frac{\alpha_{\mathrm{em}}}{\pi yQ^{2}}\left[(1-y)\left(1-\frac{Q^{2}_{\mathrm{min}}}{Q^{2}}\right)F_{1}(Q^{2})+\frac{y^{2}}{2}F_{2}(Q^{2})\right],
\displaybreak[0]\nonumber\\
\end{align}
where the specific expressions of $F_{1}(Q^{2})$ and $F_{2}(Q^{2})$ for different photon emission mechanisms have been given in Eqs. (\ref{F12p})-(\ref{L.UIC}).
Actually, the last equation of Eq. (\ref{fgamma.Gen.}) is the origin of various practically employed photon spectra~\cite{Zhu:2016zmd}, which derived by assuming that $Q^{2}_{\mathrm{min}}=y^{2}m_{\alpha}^{2}/(1-y)$, this is the leading order term of complete expression in the expansion of $\mathcal{O}(m_{\alpha}^{2})$, and is only valuable when $m_{\alpha}^{2}\ll1\ \mathrm{GeV}^2$.
However $m_{p}^{2}$ and $m_{Pb}^{2}$ do not satisfy this condition, this leads to about $10\%$ errors in various spectra.

For the case of coherent-photon emission of proton, a widely applied equivalent photon spectrum has been investigated by Kniehl~\cite{Kniehl:1990iv}, which derived from Eq. (\ref{fgamma.Gen.}) by including the effect of both the magnetic dipole moment and the corresponding magnetic form factor of the proton. By setting $Q^{2}_{\mathrm{max}}\rightarrow\infty$, he obtained the following form with $a=4m^{2}_{p}/0.71 \mathrm{GeV}^{2}=4.96$ and $b=2.79$,
\begin{eqnarray}\label{Kn.fgamma.coh.}
&&f_{\mathrm{Kn}}(y)\nonumber\\
&=&\frac{\alpha_{\mathrm{em}}}{2\pi}y\left[c_{1}x\ln\left(1+\frac{c_{2}}{z}\right)-(x+c_{3})\ln\left(1-\frac{1}{z}\right)\right.\nonumber\\
&+&\left.\frac{c_{4}}{z-1}+\frac{c_{5}x+c_{6}}{z}+\frac{c_{7}x+c_{8}}{z^{2}}+\frac{c_{9}x+c_{10}}{z^{3}}\right],
\end{eqnarray}
where $x$ and $z$ depend on $y$, $x=1/2-2/y+2/y^{2}$, $z=1+ay^{2}/4(1-y)$.

Another most important photon spectrum is the semiclassical impact parameter description, which excludes the hadronic interaction easily.
The calculation of this photon spectrum is explained in Ref.~\cite{CED}, and the final result can be presented as
\begin{align}
&f_{\mathrm{SC}}(y)\nonumber\\
=&\frac{2Z^{2}\alpha_{\mathrm{em}}}{\pi}\left(\frac{c}{\upsilon}\right)^{2}\frac{1}{y}\left[\xi K_{0}K_{1}+\frac{\xi^{2}}{2}\left(\frac{\upsilon}{c}\right)^{2}(K^{2}_{0}-K^{2}_{1})\right],\label{SC.fgamma.coh.}
\end{align}
where $\upsilon$ is the velocity of the point charge $Ze$, $K_{0}(x)$ and $K_{1}(x)$ are the modified Bessel
functions, and $\xi=b_{\textrm{min}}m_{A}y/\upsilon$.

For the case of coherent-photon emission of lead, Drees, Ellis and Zeppenfeld~\cite{Drees:1989vq} developed an equivalent photon spectrum (DEZ) which excludes the contribution of $F_{2}(Q^{2})$.
Based on the assumptions $y\ll1$, $Q^{2}_{\mathrm{max}}\rightarrow\infty$ and $F^{2}_{Pb}(Q^{2})\approx\exp(-\frac{Q^{2}}{Q^{2}_{0}})$, they obtained
\begin{eqnarray}\label{DEZ.fgamma.coh.}
&&f_{\mathrm{DEZ}}(y)\nonumber\\
&=&\!\!\frac{\alpha_{\mathrm{em}}}{\pi}\left[-\frac{\exp(-Q^{2}_{\mathrm{min}}/Q^{2}_{0})}{y}+\left(\frac{1}{y}+\frac{M^{2}}{Q^{2}_{0}}y\right)\Gamma(0,\frac{Q^{2}_{\mathrm{min}}}{Q^{2}_{0}})\right],\nonumber\\
\end{eqnarray}
where $Q^{2}_{\mathrm{min}}=m^{2}_{Pb}y^{2}$, $\Gamma(a,Q^{2}_{\mathrm{min}}/Q^{2}_{0})=\int_{y}^{\infty}t^{a-1}e^{-t}dt$.
It should be noticed that, $y\ll1$ means $Q^{2}_{\mathrm{max}}\ll1$, which contradicts with the assumption
$Q^{2}_{\mathrm{max}}\rightarrow\infty$.
This error will be discussed later on.

For the case of incoherent-photon emission of quarks, there is a widely used equivalent photon spectrum which neglects the weighting factors and takes
$Q^{2}_{\mathrm{min}}=1\ \mathrm{GeV}^{2}$ and $Q^{2}_{\mathrm{max}}=\hat{s}/4$ \cite{Yu:2015kva, Yu:2017rfi, Yu:2017pot, Fu:2011zzm, Fu:2011zzf, Chin.Phys.C_36_721},
\begin{eqnarray}\label{fgamma.incoh.}
f_{\gamma/q}&=&e_{a}^{2}\frac{\alpha_{\mathrm{em}}}{2\pi}\frac{1+(1-y)^{2}}{y}\ln\frac{Q^{2}_{\mathrm{max}}}{Q^{2}_{\mathrm{min}}}.
\end{eqnarray}

\section{NUMERICAL RESULTS}
\label{NUMERICAL RESULTS}

In this section we present the numerical results.
There are several theoretical inputs need to be provided.
The mass range of dileptons is $0.2\ \mathrm{GeV}<M<0.75\ \mathrm{GeV}$, the mass of proton is $m_{p}=0.938\ \mathrm{GeV}$~\cite{Agashe:2014kda}.
The strong coupling constant is taken as the one-loop form~\cite{Ma:2015ykd}
\begin{eqnarray}\label{alphas}
\alpha_{s}=\frac{12\pi}{(33-2n_{f})\ln(\mu^{2}/\Lambda^{2})},
\end{eqnarray}
with $n_{f}=3$ and $\Lambda=0.2\ \mathrm{GeV}$.
Furthermore, the coherence condition~\cite{Baur:2001jj} is adopted in the case of coherent-photon emission, which means that the wavelength of the photon is larger than the size of the nucleus, and the charged constituents inside the nucleus should act coherently.
This condition limits $Q^{2}$ and $y$ to very low values ($Q^{2}\leq 1/R^{2}_{A}$, $R_{A}=A^{1/3}1.2\ \mathrm{fm}$ is the size of the nucleus), $Q^{2}_{\mathrm{max}}\sim0.027\ \mathrm{GeV}^{2}$ and $7.691\times10^{-4}\ \mathrm{GeV}^{2}$, and $y_{\mathrm{max}}\sim0.16$ and $1.42\times10^{-4}$, for proton and lead respectively.
Finally, the full partonic kinematics and the bounds of involved variables are given in Appendix~\ref{FKR}.

\begin{figure*}[htbp]
  \centering
  \includegraphics[height=9cm,width=17cm]{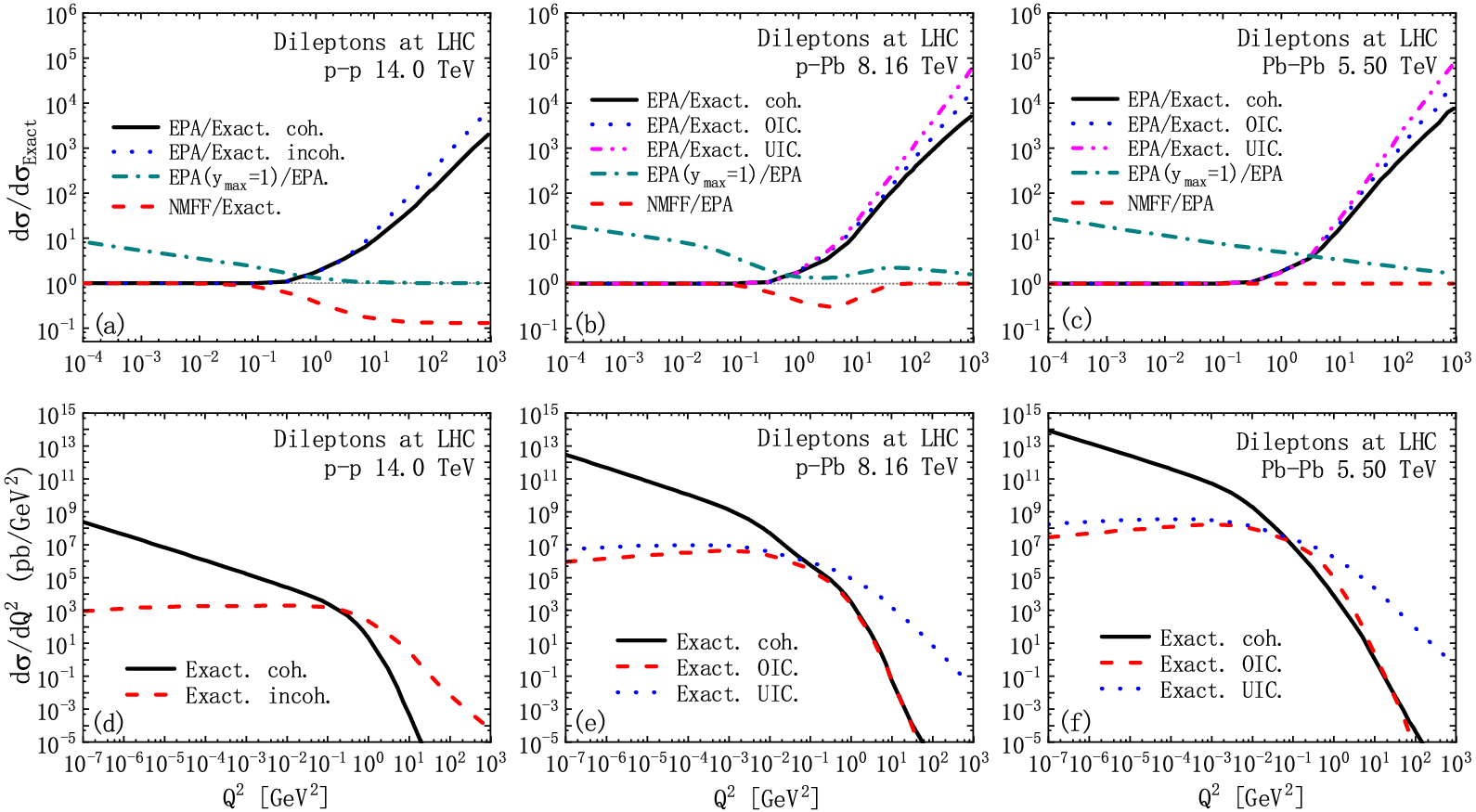}
  \caption{The $Q^{2}$ distribution of dileptons photoproduction at LHC energies.
  The upper panels show the ratios of differential cross sections in different forms to the exact ones.
  The lower panels show the exact results of $Q^{2}$ dependent differential cross sections.
  While the left, central and right panels plot the corresponding results in p-p, p-Pb and Pb-Pb collisions, respectively.
  (a)-(c): Black solid, blue dot and magenta dash dot dot lines are for the ratios of EPA result to the exact one for the coherent-photon emission[coh.(dir.+res.)], ordinary-incoherent photon emission processes [OIC.(dir.+res.)] and ultra-incoherent photon emission processes [UIC.(dir.+res.)], respectively.
  Red dash line---the ratio of the result with no contribution of magnetic form factor (NMFF) to the exact one.
  Dark cyan dash dot line---the ratio of EPA result with $y_{\mathrm{max}}=1$ to the exact one.
  }
  \label{Q2.dileptons}
\end{figure*}

\begin{figure*}[htbp]
  \includegraphics[height=9cm,width=17cm]{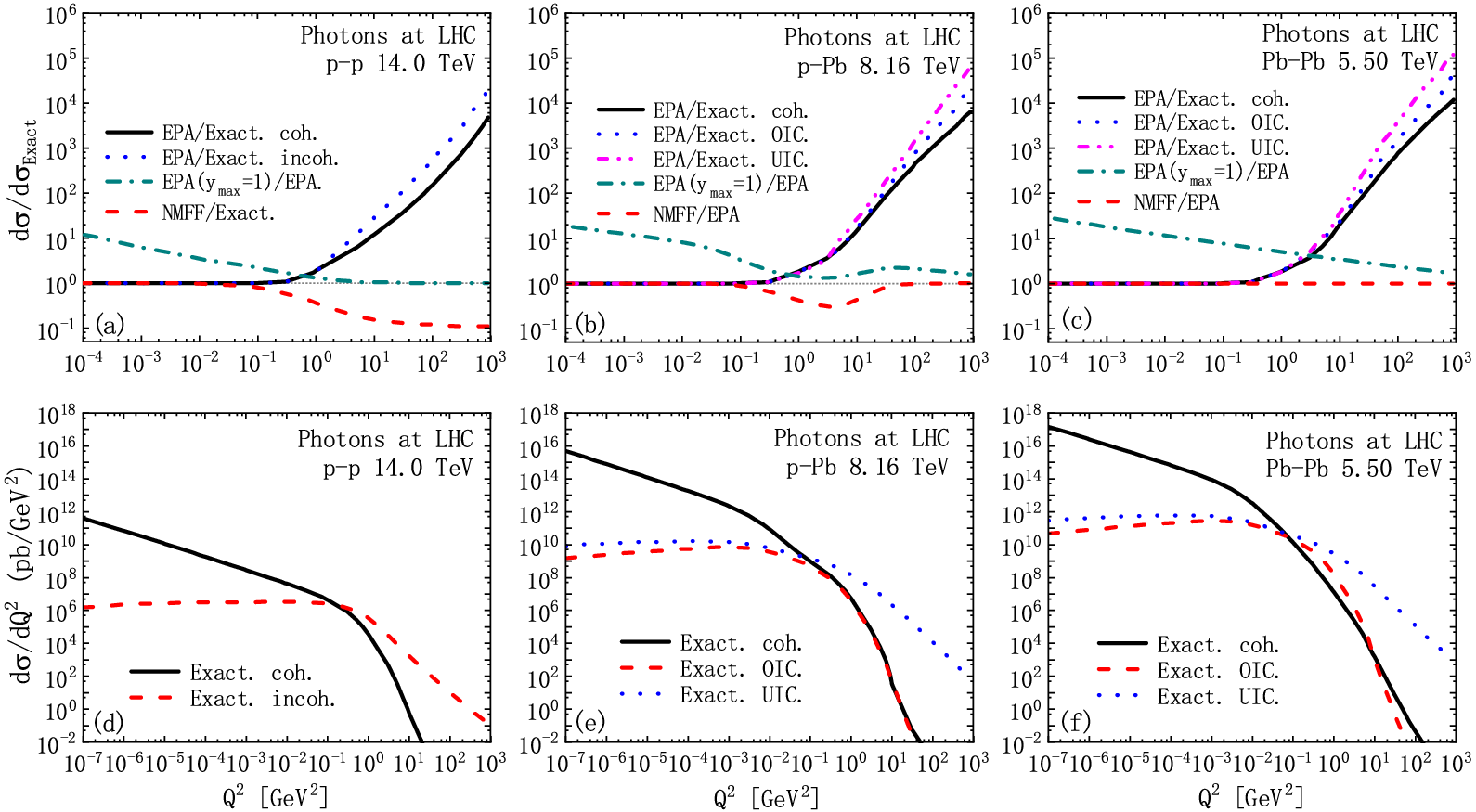}
  \caption{Same as Fig. \ref{Q2.dileptons} but for photons photoproduction.
  }
  \label{Q2.photons}
\end{figure*}

\begin{figure*}[htbp]
  \includegraphics[height=9cm,width=17cm]{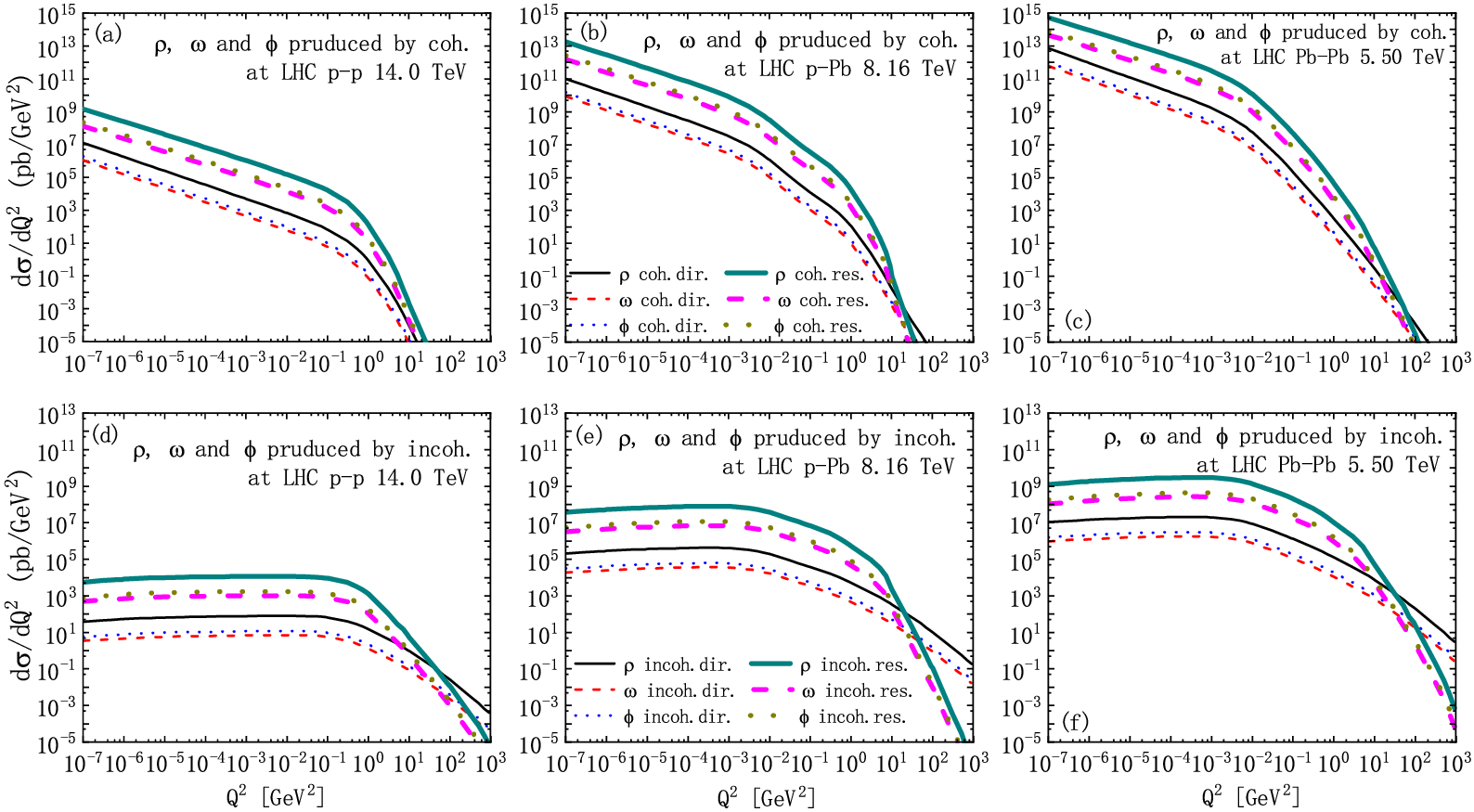}
  \caption{The exact results of $Q^{2}$ dependent differential cross sections for light vector mesons photoproduction.
  The left, central and right panels show the differential cross sections in p-p, p-Pb and Pb-Pb collisions, respectively.
  The upper and lower panels plot the differential cross section in the channel of coherent-photon emission and incoherent-photon emission, respectively.
  The curves in panels (e) and (f) are the sum of OIC. and UIC.
  }
  \label{Q2.LVM}
\end{figure*}

In Fig.~\ref{Q2.dileptons}, the upper panels show the ratios of differential cross sections in different forms to the exact ones for the dileptons photoproduction.
The lower panels show the exact results of $Q^{2}$ dependent differential cross sections.
While the corresponding results in p-p, p-Pb and Pb-Pb collisions are presented in left, central and right panels.
In upper panels [Fig.~\ref{Q2.dileptons}(a)-(c)], the EPA results are almost the same as the exact ones in small $Q^{2}$ region, the differences appear when $Q^{2}>0.1\ \mathrm{GeV}^{2}$ and become evident at large values of $Q^{2}$.
These differences are larger and largest in the channels of ordinary-incoherent and ultra-incoherent photon emissions, and become more and more obvious
in p-Pb and Pb-Pb collisions, respectively.
On the contrary, the ratios of the results with $y_{\mathrm{max}}=1$ to the EPA ones are largest in small $Q^{2}$ region, and decrease with increasing $Q^{2}$.
Comparing with the case of p-p collisions, these ratios are larger and exist in the whole $Q^{2}$ region in p-Pb and Pb-Pb collisions
(at $Q^{2}=10^{-4}\ \mathrm{GeV}^{2}$, the ratios are $8.3$, $19.1$ and $28.4$ in p-p, p-Pb and Pb-Pb collisions, respectively).
Therefore, EPA is only applicable in very restricted domain (small $y$ and $Q^{2}$ domains), its errors appear when $y>0.29$ and $Q^{2}>0.1~\textrm{GeV}^{2}$, and become prominent in p-Pb and Pb-Pb collisions.

We find that in panel (a), the result without the effect of magnetic form factor (NMFF) nicely agrees with exact one when $Q^{2}<0.05\ \mathrm{GeV}^{2}$, the difference becomes evident when $Q^{2}>1\ \mathrm{GeV}^{2}$.
Therefore, the contribution of magnetic form factor is concentrate on the large $Q^{2}$ domain.
However in panel (b), the difference only appears at $0.05\ \mathrm{GeV}^{2}< Q^{2}< 100\ \mathrm{GeV}^{2}$, where the curves are distorted.
Since in p-Pb collision, the photon emitter can be both proton ($\gamma Pb$) and lead ($\gamma p$).
The distortion is caused by the contribution of $\gamma Pb$, and the difference also comes from the proton magnetic form factor $G_{\mathrm{M}}(Q^{2})$ in $\gamma Pb$ process.
Thus, the process $\gamma Pb$, which is usually neglected in p-Pb collision ~\cite{Baltz:2007kq}, has non-negligible effect in large $Q^{2}$ region.
In panel (c) we observe that NMFF is consistent with the exact one in the whole $Q^{2}$ region,
since the effect of magnetic form factor of lead can be neglected compared to its electric form factor which is enhanced by the factor $Z_{Pb}^{2}$.

In the lower panels [Fig.~\ref{Q2.dileptons}(d)-(f)], the coherent and incoherent reactions dominant the small and large $Q^{2}$ regions, respectively.
They become comparable at $Q^{2}=0.1\ \mathrm{GeV}^{2}$.
Comparing with the features of EPA derived from upper panels, one can see that EPA is a good approximation for coherent and ordinary-incoherent reactions.
However, EPA is in contradiction with ultra-incoherent reactions and will cause the significant errors which become rather serious in p-Pb and Pb-Pb collisions.
On the other hand, we also find that the contributions of ultra-incoherent photon emissions are always much larger than those of ordinary-incoherent photon emissions in the whole $Q^{2}$ domain, we will discuss this point quantitatively in the following Tables.

Fig.~\ref{Q2.photons} is similar to Fig.~\ref{Q2.dileptons} but for photons photoproduction, where the ratios are much more evident.
Finally, we calculate the exact results of light vector mesons photoproduction in Fig.~\ref{Q2.LVM}.
We find that the revolved contributions are generally two orders of magnitudes (OOMs) larger than direct contributions, thus the EPA errors in photoproduction processes are mainly come from the resolved contributions.

\begin{table}[htbp]\small
\renewcommand\arraystretch{1.7}
\setlength{\abovecaptionskip}{1mm}
\centering
\caption{\label{Total.CS.coh.} Total cross sections of the photons photoproduction in the channel of coherent-photon emissions [coh.(dir.+res.)].}
\begin{tabular}{L{1.7cm}C{0.8cm}C{2.1cm}L{1.8cm}c}
\hline
\hline
coh.                                                        & Exact   & EPA CC\footnotemark[2]     & \makecell[l]{EPA\\ \scriptsize($Q^{2}_{\mathrm{max}}\backsim \hat{s}$)} & \makecell[l]{EPA \\ \footnotesize ($y_{\mathrm{max}}=1$)} \\
\hline
\footnotesize $\sigma_{\mathrm{pp}}$~$[\mathrm{nb}]$        & 70.35   & 70.37  & 110.14    & 1200.48   \\
\footnotesize $\delta_{\mathrm{pp}}$~$[\%]$                 & 0.0     & 0.03   & 56.56     & 1606.50   \\
\hline
\footnotesize $\sigma_{\mathrm{pPb}}$~$[\mu\mathrm{b}]$     & 357.26  & 357.25 & 581.55    & 17949.33  \\
\footnotesize $\delta_{\mathrm{pPb}}$~$[\%]$                & 0.0     & 0.0    & 62.78     & 4924.22   \\
\hline
\footnotesize $\sigma_{\mathrm{PbPb}}$~$[\mathrm{mb}]$      & 13.45   & 13.45  & 23.01     & 1318.02   \\
\footnotesize $\delta_{\mathrm{PbPb}}$~$[\%]$               & 0.0     & 0.0    & 71.05     & 9699.04   \\
\hline
\hline
\end{tabular}
\footnotetext[1]{ Relative error to the exact result: $\delta=\sigma/\sigma_{\textrm{Exact}}-1$.}
\footnotetext[2]{ EPA result with the coherence condition (CC).}
\end{table}

To quantitatively estimate the errors caused by the widely adopted kinematical limitations, and to discuss the double counting encountered in literatures~\cite{Drees:1989vq, Drees:1988pp, Drees:1994zx, Frixione:1993yw, Zhu:2015via, Zhu:2015qoz, Fu:2011zzm, Fu:2011zzf, Chin.Phys.C_36_721, Yu:2017rfi, Yu:2015kva, Yu:2017pot, Nystrand:2004vn, Nystrand:2006gi,  Kniehl:2001tk, Kniehl:1990iv, sp}, we calculate the total cross sections in Table~\ref{Total.CS.coh.}-\ref{Total.CS.UIC.}.
In Table~\ref{Total.CS.coh.}, the relative errors caused by $Q^{2}_{\mathrm{max}}\sim \hat{s}$ are evident, but those caused by $y_{\mathrm{max}}=1$ are rather serious.
These relative errors gradually increase from p-p to Pb-Pb collisions.
However, the EPA results with coherence condition (CC) nicely agree with exact ones, since CC limits $y_{\mathrm{max}}$ and $Q^{2}_{\mathrm{max}}$ to very low values which effectively avoid the errors from large $y$ and $Q^{2}$ domains.
Therefore, EPA is very sensitive to the values of $y_{\textrm{max}}$ and $Q^{2}_{\textrm{max}}$, the common options~$Q^{2}_{\mathrm{max}}\sim \hat{s}$ or even $\infty$, and $y_{\mathrm{max}}=1$ will cause the large errors.

\begin{table}[htbp]\small
\renewcommand\arraystretch{1.7}
\setlength{\abovecaptionskip}{1mm}
\centering
\caption{\label{Total.CS.OIC.} Same as Table. \ref{Total.CS.coh.} but in the channel of ordinary-incoherent photon emissions [OIC.(dir.+res.)].}
\begin{tabular}{L{1.6cm}L{1cm}L{1.1cm}L{1.5cm}L{1.5cm}C{1cm}}
\hline
\hline
OIC.                            & Exact   & \makecell[l]{EPA \\ CC} & \makecell[l]{EPA \\ \scriptsize($Q^{2}_{\mathrm{max}}\backsim \hat{s}$)} & \makecell[l]{EPA \\ \footnotesize ($y_{\mathrm{max}}=1$)}         & \makecell[l]{EPA\\ no WF}\\
\hline
\footnotesize $\sigma_{\mathrm{pPb}}$ [$\mu\mathrm{b}$]         & 2.78   & 2.77   & 6.73    & 9.67    & 7.97   \\
\footnotesize $\delta_{\mathrm{pPb}}~[$\%$]$                    & 0.0    & 0.0    & 142.33  & 248.37  & 186.98 \\
\hline
\footnotesize $\sigma_{\mathrm{PbPb}}$ [$\mu\mathrm{b}$]        & 117.79 & 117.84 & 288.80  & 495.42  & 352.81 \\
\footnotesize $\delta_{\mathrm{PbPb}}~[$\%$]$                   & 0.0    & 0.04   & 145.18  & 320.59  & 199.52 \\
\hline
\hline
\end{tabular}
\end{table}

\begin{table}[htbp]\small
\renewcommand\arraystretch{1.75}
\setlength{\abovecaptionskip}{0mm}
\centering
\caption{\label{Total.CS.UIC.} Same as Table. \ref{Total.CS.coh.} but in the channel of ultra-incoherent photon emissions [UIC.(dir.+res.)].}
\begin{tabular}{L{1.2cm}C{1.5cm}cC{2cm}c}
\hline
\hline
UIC.                                                        & Exact   & EPA      & \makecell[l]{EPA \footnotesize no WF} & \makecell[l]{EPA \footnotesize no WF\\ \footnotesize ($Q^{2}_{\mathrm{min}}=1\mathrm{GeV}$)} \\
\hline
\footnotesize $\sigma_{\mathrm{pp}}$~$[\mathrm{nb}]$        & 62.55   & 292.04   & 472.52    & 260.97  \\
\footnotesize $\delta_{\mathrm{pp}}$~$[\%]$                 & 0.0     & 366.85   & 655.38    & 317.18  \\
\hline
\footnotesize $\sigma_{\mathrm{pPb}}$~$[\mu\mathrm{b}]$     & 22.27   & 174.40   & 227.23    & 142.80  \\
\footnotesize $\delta_{\mathrm{pPb}}$~$[\%]$                & 0.0     & 683.24   & 920.49    & 541.32  \\
\hline
\footnotesize $\sigma_{\mathrm{PbPb}}$~$[\mu\mathrm{b}]$    & 812.43  & 8219.03  & 9659.63   & 6116.08 \\
\footnotesize $\delta_{\mathrm{PbPb}}$~$[\%]$               & 0.0     & 911.66   & 1088.98   & 653.16  \\
\hline
\hline
\end{tabular}
\end{table}

Table~\ref{Total.CS.OIC.} is similar to Table~\ref{Total.CS.coh.} but for the case of ordinary-incoherent photon emissions, where the relative errors are still evident.
In Table~\ref{Total.CS.UIC.}, we can quantitatively check the inapplicability of EPA in ultra-incoherent reactions, where the EPA errors are prominent and become larger and largest in p-Pb and Pb-Pb collisions.
We find that the EPA results without weighting factor are the nonsense large values, these unphysical results are caused by the double counting which are much more serious in ultra-incoherent reactions compared to ordinary-incoherent reactions in Table~\ref{Total.CS.OIC.}.
However, this trouble is often neglected in the most works~\cite{Zhu:2015qoz, Fu:2011zzm, Fu:2011zzf, Chin.Phys.C_36_721, Yu:2015kva, Yu:2017rfi, Yu:2017pot} where an artificial cutoff $Q^{2}>1\ \mathrm{GeV}^{2}$ is adopted, but we can see that in Table~\ref{Total.CS.UIC.} the corresponding results are still not accurate.
Thus, the weighting factor can effectively and naturally avoid double counting, and the exact treatment is needed for ultra-incoherent photon emissions.
Otherwise, we observe that the exact results of ultra-incoherent photon emissions in Table~\ref{Total.CS.UIC.} are much larger than those of ordinary-incoherent photon emissions in Table~\ref{Total.CS.OIC.}, and it is even comparable with coherent-photon emissions in p-p collisions in Table~\ref{Total.CS.coh.}.
Thus, ultra-incoherent photon emissions is the important channel of photoproduction processes, especially when $Z$ is not much larger than one.

%

\begin{figure*}[htbp]
  \includegraphics[height=12cm,width=16cm]{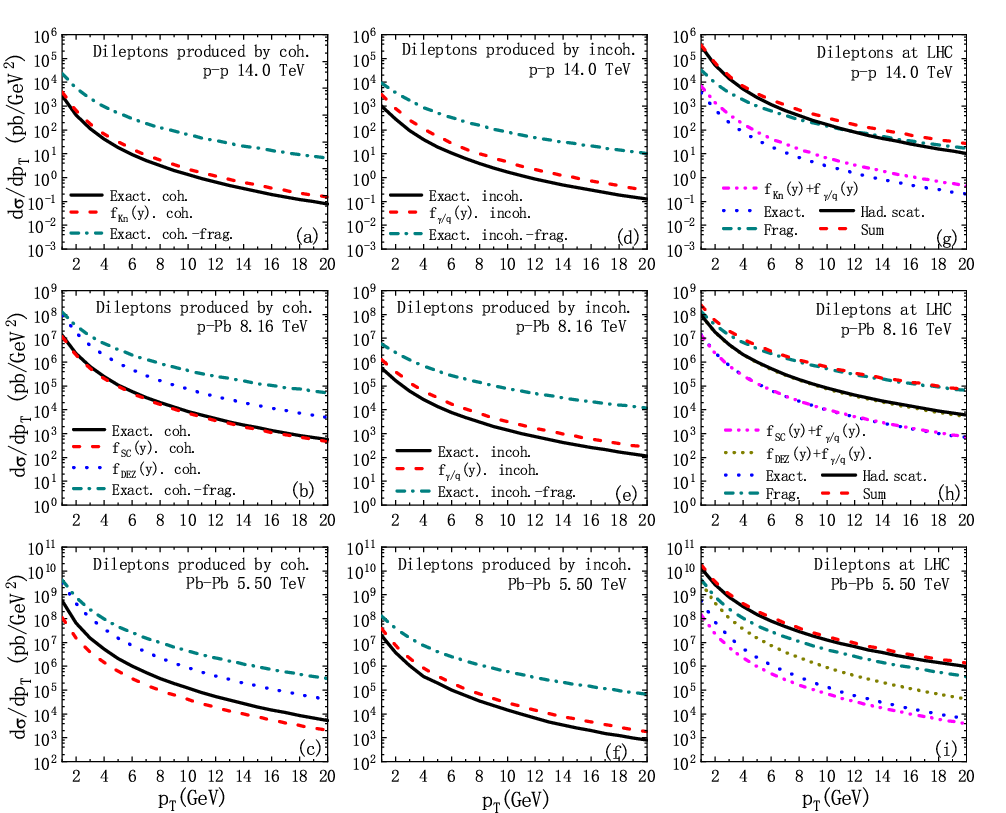}
  \caption{The $p_{T}$ distribution of dileptons photoproduction at LHC energies.
  The left and central panels show the differential cross sections in the channels of coherent-photon emission [coh.(dir.+res.)] and incoherent-photon emission processes [incoh.(dir.+res.)], respectively.
  The right panels show the comparisons between the photoproduction processes and the hard scattering of initial partons (had.scat.).
  While the upper, central and lower panels plot the corresponding results in p-p, p-Pb and Pb-Pb collisions, respectively.
  (g)-(i): Black solid line is for had.scat., blue dot line is for the exact results of photoproduction processes (coh.+incoh.), dark cyan dash dot line denotes the exact result of fragmentation dileptons photoproduction [(coh.+incoh.)-frag.], magenta dash dot dot line and yellow short dash line denote the EPA results based on the equivalent photon spectra.
  Red dash line is for the sum of had.scat., direct dileptons and fragmentation dileptons photoproductions.
  It should be emphasized that the curves in panels (e) and (f) are the sum of OIC. and UIC.
}
\label{pT.dileptons}
\end{figure*}

\begin{figure*}[htbp]
\includegraphics[height=12cm,width=16cm]{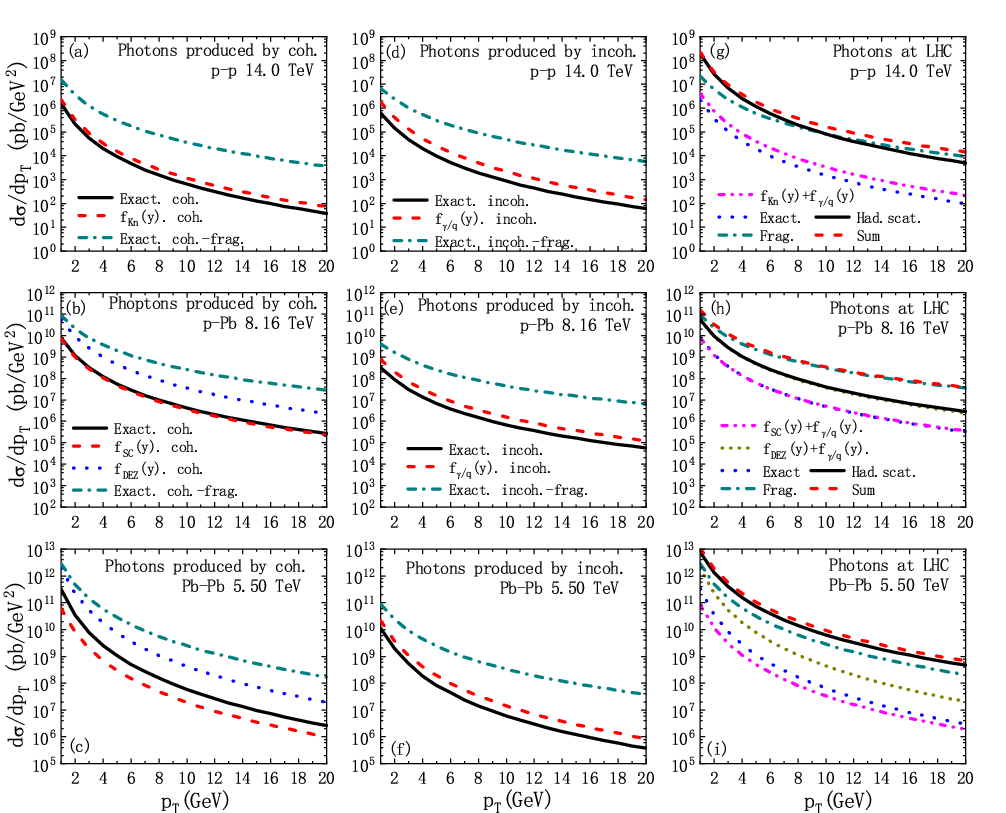}
\caption{Same as Fig. \ref{pT.dileptons} but for photons production.}
\label{pT.photons}
\end{figure*}

\begin{figure*}[htbp]
\includegraphics[height=12cm,width=16cm]{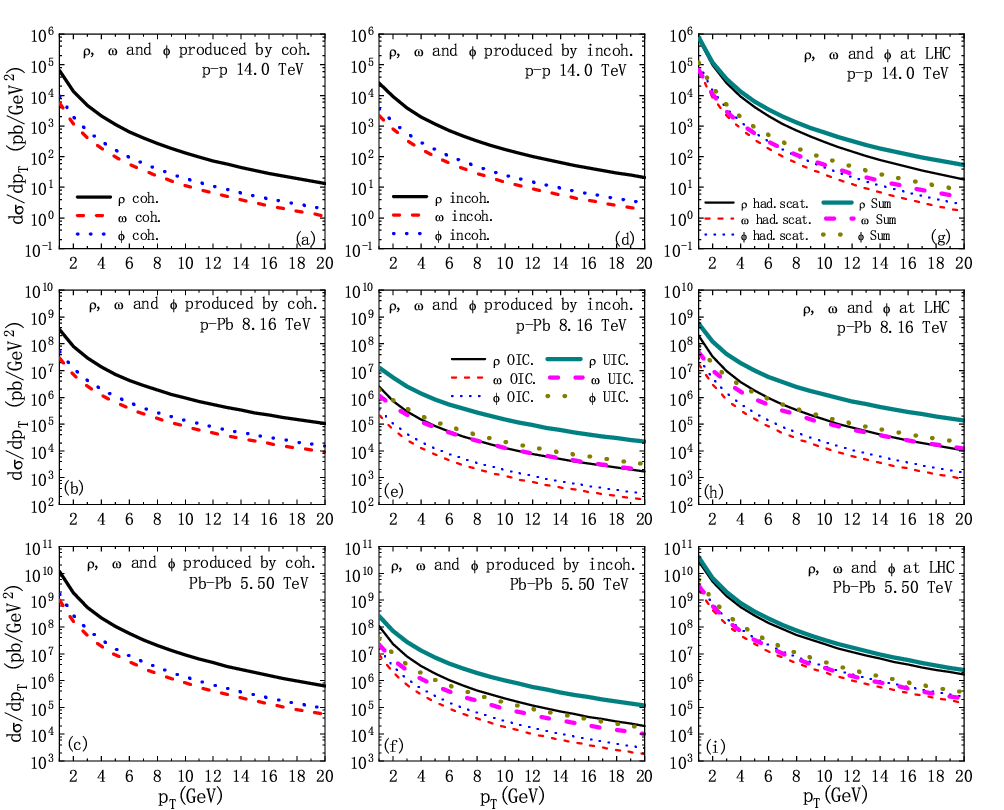}
\caption{The exact results of $p_{T}$ dependent differential cross sections for light vector mesons photoproduction.
  The left and cental panels show the differential cross sections in the channels of coherent-photon emission [coh.(dir.+res.)] and incoherent-photon emission processes [incoh.(dir.+res.)], respectively.
  The right panels show the comparisons between the photoproduction processes and the hard scattering of initial partons.
  While the upper, central and lower panels plot the corresponding results in p-p, p-Pb and Pb-Pb collisions, respectively.
  In panels (g)-(i), the black solid, red dash and blue dot lines are for had.scat. of $\rho$, $\omega$ and $\phi$, respectively.
  While those thick one with different colors are the sum of had.scat. and photoproduction processes.
}
\label{pT.LVM}
\end{figure*}

In order to discuss the features of the photon spectra which are widely employed in most works, and estimate the contribution of photoproduction processes.
We plot the $p_{T}$ dependent differential cross sections of dileptons production in Fig.~\ref{pT.dileptons},
the left and central panels show the comparisons between the results based on the spectra mentioned in Sec.\ref{EPASp} and the exact ones in the channels of coherent- [coh.(dir.+res.)] and incoherent-photon emissions [incoh.(dir.+res.)], respectively.
The right panels show the comparisons between the photoproduction processes and the initial partons hard scattering (had.scat.).
While the upper, central and lower panels plot the corresponding results in p-p, p-Pb and Pb-Pb collisions.
In Fig.~\ref{pT.dileptons} (a)-(f), the results based on the referred spectra generally have the non-negligible deviations from the exact ones.
The spectrum $f_{\mathrm{Kn}}$ [Eq.~(\ref{Kn.fgamma.coh.})] adopts $Q^{2}_{\mathrm{max}}=\infty$ and $y_{\mathrm{max}}=1$ which include the large EPA errors from large $Q^{2}$ and $y$ domains.
Besides, $f_{\mathrm{Kn}}$ includes the effect of magnetic form factor which is concentrate on the large $Q^{2}$ domain and should essentially be excluded.
The errors caused by $f_{\mathrm{DEZ}}$ [Eq.~(\ref{DEZ.fgamma.coh.})] are largest, since $f_{\mathrm{DEZ}}$ is based on the assumptions, $Q^{2}_{\mathrm{max}}\sim\infty$ and $y\ll1$, which are contradict with each other ($Q^{2}_{\mathrm{max}}\sim\infty$ means $y_{\mathrm{max}}=1$).
One exception is the result of $f_{\mathrm{SC}}$ [Eq.~(\ref{SC.fgamma.coh.})] which nicely agrees with the exact one in p-Pb collision, since this semiclassical photon flux effectively excludes the hadronic interactions.
But its deviation still can not be neglected in Pb-Pb collision.
Finally, the results of incoherent photon spectrum $f_{\gamma/q}$ [Eq.~(\ref{fgamma.incoh.})] are about five times larger than the exact ones in each case,  this verifies again the inapplicability of EPA for ultra-incoherent reactions.
Actually, the errors of $f_{\gamma/q}$ should be much larger, but an artificial cutoff $Q^{2}_{\mathrm{min}}=1\ \mathrm{GeV}^{2}$ is employed for avoiding the unphysical large value caused by double counting.

In Fig.~\ref{pT.dileptons} (g)-(i), we observe that the contributions of fragmentation dileptons production are generally about one and two OOMs larger than those of direct dileptons production in small and large $p_{T}$ domains, respectively.
It is even larger than had.scat. when $p_{T} > 10\ \mathrm{GeV}$ in p-p collision [panel (g)], and is an OOM larger than had.scat. in the whole $p_{T}$ region in p-Pb collision [panel (h)].
Hence, fragmentation processes dominate the photoproduction processes at LHC energies.
On the other hand, we find that photoproduction processes give the non-negligible corrections to had.scat. in p-p and Pb-Pb collisions, especially in the large $p_{T}$ domain.
And in p-Pb collision, photoproduction processes start to play the fundamental role in the production of dileptons.
One can see that the results of equivalent photon spectra provide the large fictitious contributions to dileptons production, and thus the results in Refs.~\cite{Zhu:2015qoz, Yu:2017rfi, Yu:2017pot, Yu:2015kva, Fu:2011zzm,Fu:2011zzf,Chin.Phys.C_36_721} are not accurate enough, where the referred equivalent photon spectra are adopted and the serious double counting exists.
In addition, we also find that the contributions of coherent-photon emission are much larger than those of ordinary-incoherent and ultra-incoherent photon emissions in p-Pb and Pb-Pb collisions, since its enhanced by $Z_{Pb}^{2}$.
However, this is very different from the results in Ref.~\cite{Yu:2015kva} where the situation is opposite.

Fig~\ref{pT.photons} is similar to Fig~\ref{pT.dileptons} but for photons photoproduction.
The errors of referred spectra and the contribution of photoproduction processes are more obvious.
Finally, we calculate the exact results of light vector mesons photoproduction in Fig.~\ref{pT.LVM}.
We observe that in panels (e) and (f) the contributions of ultra-incoherent photon emissions are an OOM larger than those of ordinary-incoherent photon emissions,
this verifies again the views derived from Table~\ref{Total.CS.UIC.}, that ultra-incoherent photon emission is the important channel of photoproduction processes in heavy-ion collisions, which should not be neglected in the calculations.

\begin{figure*}[htbp]
\setlength{\abovecaptionskip}{0.mm}
\includegraphics[height=11cm,width=15cm]{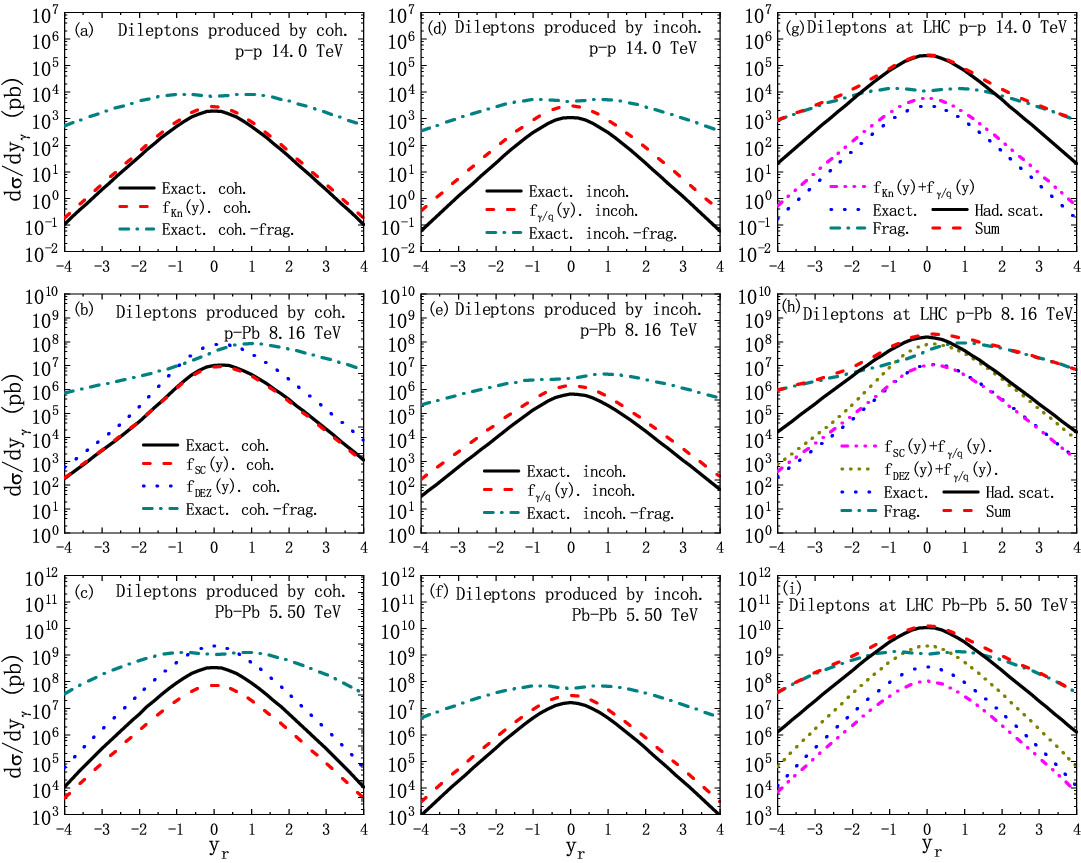}
\caption{Same as Fig. \ref{pT.dileptons} but for $y_{r}$ distribution.
}
\label{yr.dileptons}
\end{figure*}

\begin{figure*}[htbp]
\setlength{\abovecaptionskip}{0.mm}
\includegraphics[height=11cm,width=15cm]{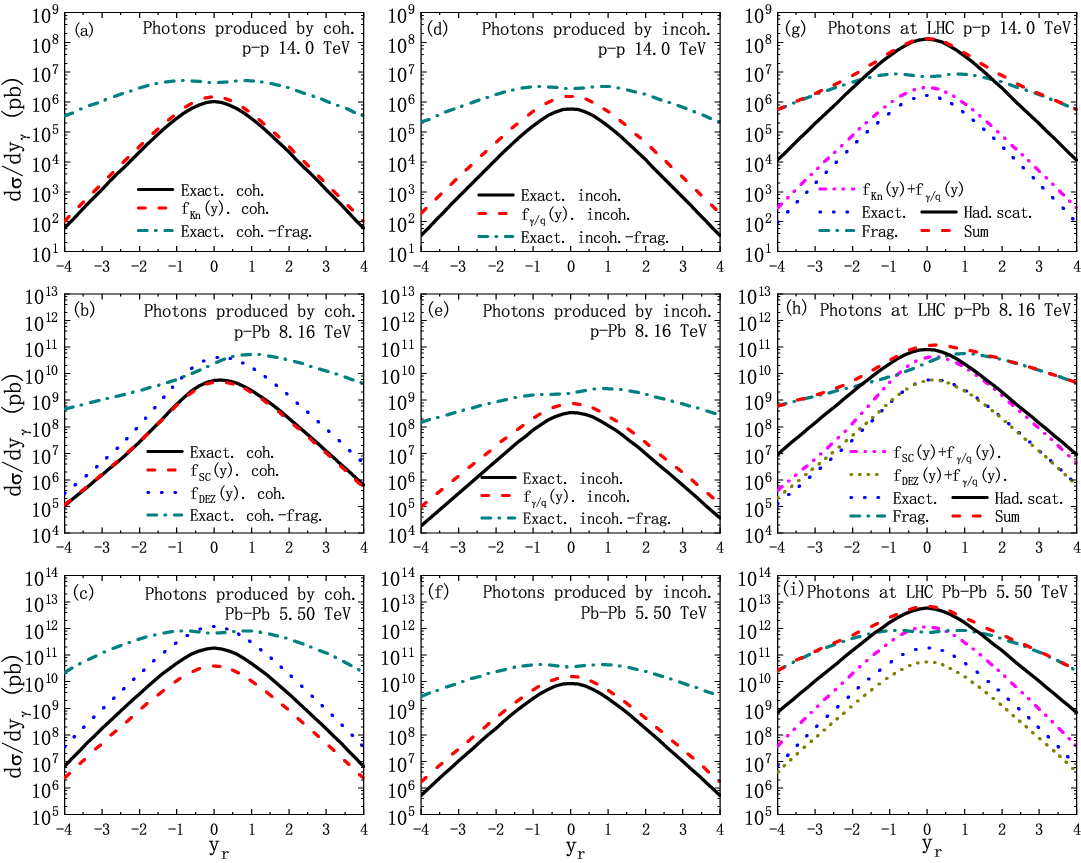}
\caption{Same as Fig. \ref{pT.photons} but for $y_{r}$ distribution.
}
\label{yr.photons}
\end{figure*}

\begin{figure*}[htbp]
\setlength{\abovecaptionskip}{0.mm}
\includegraphics[height=11cm,width=15cm]{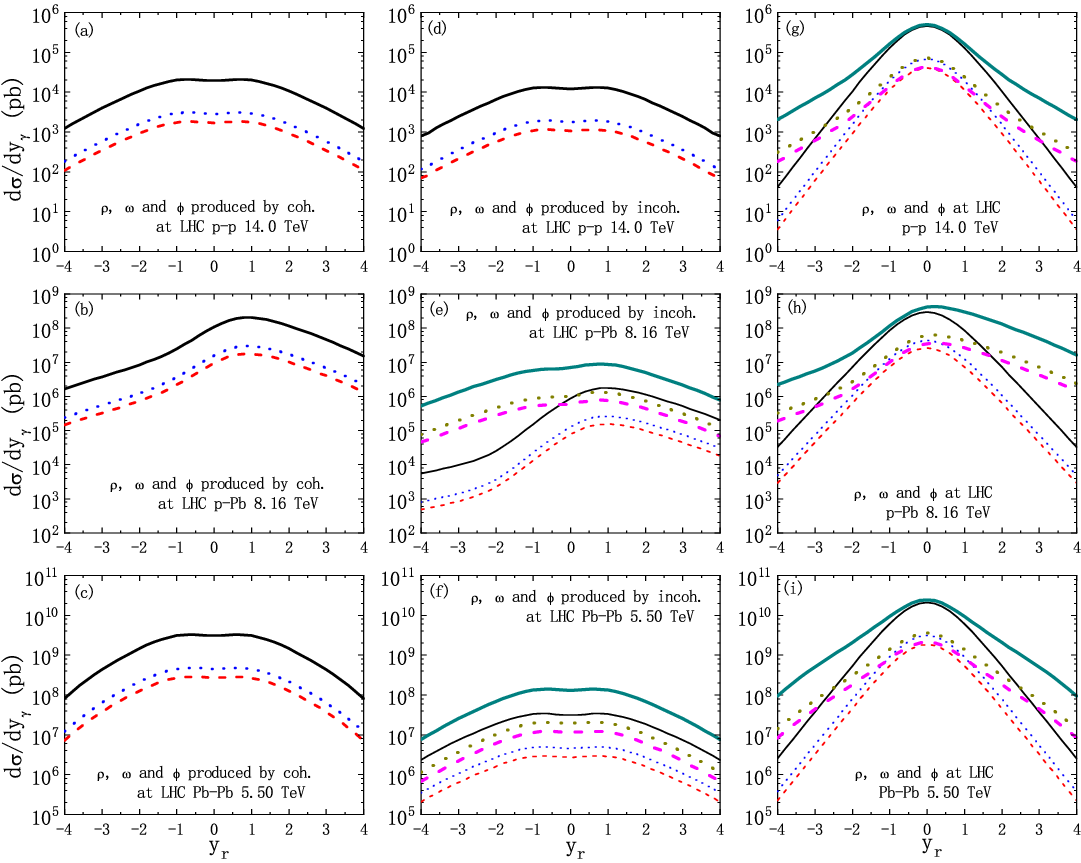}
\caption{Same as Fig. \ref{pT.LVM} but for $y_{r}$ distribution.
}
\label{yr.LVM}
\end{figure*}

In Figs.~\ref{yr.dileptons}-\ref{yr.LVM}, the $y_{r}$ distributions are plotted.
It can be seen that the contributions are dominant in the central $y_{r}$ region.
The EPA results based on the referred photon flux functions generally have non-negligible errors compared to the exact ones in the whole $y_{r}$ region.
Besides, the contributions of fragmentation dileptons photoproduction are an OOM larger than those of direct dileptons photoproduction in central $|y_{r}|$ region, and become four OOMs larger in large values of $y_{r}$ [panels (a)-(f)].
And in panels (g)-(i), we can see that the photoproduction processes give the evident corrections to had.scat., especially in the large $y_{r}$ domain.

\section{SUMMARY AND CONCLUSION}
\label{Sum.con}

In this work, we studied the photoproduction of large $p_{T}$ dileptons, photons and light vector mesons in p-p, p-Pb and Pb-Pb collisions at LHC energies.
An exact treatment which recovers the EPA in the limit $Q^{2}\rightarrow0$, is developed by performing a consistent analysis of the terms neglected in going from the accurate expression to the EPA one, in which the density of virtual photon is expanded by using the transverse and longitudinal polarization operators, and the square of electric form factor is used as weighting factor for avoiding double counting.
And the full kinematical relations are also achieved.
In order to derive in details EPA in heavy-ion collisions, we expressed the comparisons between the EPA results and the exact ones as the distribution in $Q^{2}$.
To quantitatively estimate the errors caused by the common options of kinematical limitations, and to discuss the double counting encountered in most works~\cite{Zhu:2015qoz, Fu:2011zzm, Fu:2011zzf, Chin.Phys.C_36_721, Yu:2015kva, Yu:2017rfi, Yu:2017pot}, we calculated the total cross sections.
In the sequel, we plotted the $p_{T}$ and $y_{r}$ dependent differential cross sections to estimate the contribution of photoproduction processes and to discuss the features of the photon spectra which are widely employed in most works.

The numerical results indicate that the contribution of photoproduction processes is evident in the large $p_{T}$ and $y_{r}$ domains, which is mainly come from the fragmentation processes.
And in p-Pb collisions the photoproduction processes start to play the fundamental role which is larger than had.scat. in the whole $p_{T}$ region.
Otherwise, the ultra-incoherent photon channel provides meaningful contributions to the photoproduction processes, especially when $Z$ is not much larger than 1.

On the other hand, EPA is only applicable in the small $y$ and $Q^{2}$ regions, and is very sensitive to the values of $y_{\mathrm{max}}$ and $Q^{2}_{\mathrm{max}}$.
The EPA errors appear when $y>0.29$ and $Q^{2}>0.1~\mathrm{GeV}^{2}$, and become larger and largest in p-Pb and Pb-Pb collisions.
The common options $y_{\mathrm{max}}=1$ and $Q^{2}_{\mathrm{max}}\sim\hat{s}$ or $\infty$ will cause the large errors.
These features are compatible with coherent and ordinary-incoherent reactions, but is essentially in contradiction with ultra-incoherent reactions and will cause the significant errors which become rather serious in p-Pb and Pb-Pb collisions.
Furthermore, the serious double counting exist when the different photon emission mechanisms are considered simultaneously.
And the several widely used equivalent photon spectra generally lead non-negligible errors, and the statements in literatures~\cite{Drees:1989vq, Drees:1988pp, Frixione:1993yw, Zhu:2015via, Zhu:2015qoz, Fu:2011zzm, Fu:2011zzf, Chin.Phys.C_36_721, Yu:2015kva, Yu:2017rfi, Yu:2017pot, Nystrand:2004vn, Nystrand:2006gi} are imprecise.
Therefore, the exact treatment needs to be adopted when dealing with widely kinematical regions.

\section*{ACKNOWLEDGMENTS}

We thank Dr. Yong-Ping Fu at Dianxi Science and Technology Normal University and Prof. Joakim  Nystrand at University of Bergen for useful communications.
This work is supported in part by National Key R \& D Program of China under grant No. 2018YFA0404204, the NSFC (China) grant Nos. 11747086 and 11575043, and by the Young Backbone Teacher Training Program of Yunnan University.
Z. M. is supported by the fellowship of China Postdoctoral Science Foundation under grant No. 2021M692729, and by Yunnan Provincial New Academic Researcher Award for Doctoral Candidates.

\appendix
\section{Full kinematical relations}
\label{FKR}

We give here, for completeness and the reader's convenience, a detailed account of the partonic kinematics which is matched with the exact treatment in Section~\ref{sec:GeneralF_ET}.

The energy and momentum in $\alpha b$ CM frame read
\begin{align}\label{Mant.coh.dir}
&E_{\alpha}=\frac{1}{2\sqrt{s_{0}}}(s_{0}+m_{\alpha}^{2}-m_{b}^{2}),\displaybreak[0]\nonumber\\
&E_{b}=\frac{1}{2\sqrt{s_{0}}}(s_{0}+m_{b}^{2}-m_{\alpha}^{2}),\displaybreak[0]\nonumber\\
&p_{\mathrm{CM}}=\frac{1}{2\sqrt{s_{0}}}\sqrt{(s_{0}-m_{\alpha}^{2}-m_{b}^{2})^{2}-4m_{\alpha}^{2}m_{b}^{2}},\displaybreak[0]
\end{align}
where $s_{0}=(p_{\alpha}+p_{b})^{2}$ is the CM energy square, its specific expressions for each photon emission processes are
\begin{align}\label{s0}
  &s_{0}|_{\mathrm{coh}.}=m_{A}^{2}+m_{b}^{2}+\frac{x_{b}}{N_{B}}(s-m_{A}^{2}-m_{B}^{2}),\displaybreak[0]\nonumber\\
  &s_{0}|_{\mathrm{OIC}.}=m_{p}^{2}+m_{b}^{2}+\frac{x_{b}}{N_{A}N_{B}}(s-m_{A}^{2}-m_{B}^{2}),\displaybreak[0]\nonumber\\
  &s_{0}|_{\mathrm{UIC}.}=m_{q}^{2}+m_{b}^{2}+\frac{x_{a}x_{b}}{N_{A}N_{B}}(s-m_{A}^{2}-m_{B}^{2}),\displaybreak[0]
\end{align}
where $s=(p_{A}+p_{B})^2=(N_{A}+N_{B})^{2}s_{NN}/4$ is the energy square of $AB$ CM frame.

For the case of direct photoproduction processes, the involved Mandelstam variables are given by
\begin{align}\label{Mant.}
\hat{s}&=(q+p_{b})^{2}=y(s_{0}-m_{\alpha}^{2}-m_{b}^{2})+m_{b}^{2}-Q^{2},\displaybreak[0]\nonumber\\
\hat{t}&=(q-p_{c})^{2}=(z_{q}-1)(\hat{s}+Q^{2}),\displaybreak[0]\nonumber\\
\hat{u}&=(p_{b}-p_{c})^{2}=M^{2}-z_{q}(\hat{s}+Q^{2}),\displaybreak[0]
\end{align}
where $z_{q}=(p_{c}\cdot p_{b})/(q\cdot p_{b})$ is the inelasticity variable.
For the case of resolved photoproduction processes, $\hat{s}$ in Eq. (\ref{Mant.}) should be changed as $\hat{s}_{\gamma}=(p_{a'}+p_{b})^{2}=yz_{a'}(s_{0}
-m_{\alpha}^{2}-m_{b}^{2})+m_{a'}^{2}+m_{b}^{2}$, and $z_{q}=(p_{c}\cdot p_{b})/(p_{a'}\cdot p_{b})$.

In the $p_{T}$ and $y_{r}$ distributions, the detailed expression of the Jacobian determinant $\mathcal{J}$ is
\begin{eqnarray}\label{Jac}
\mathcal{J}_{\mathrm{coh.dir.}}
=\frac{N_{B}\sqrt{(s+Q^{2}-m_{b}^{2})^{2}+4Q^{2}m_{b}^{2}}}{y(s-m_{A}^{2}-m_{B}^{2})(1-\cosh y_{r}m_{T}/\sqrt{\hat{s}})},
\end{eqnarray}
and $\mathcal{J}$ for the rest cases are: $\mathcal{J}_{\mathrm{OIC.dir.}}=N_{A}\mathcal{J}_{\mathrm{coh.dir.}}$, $\mathcal{J}_{\mathrm{UIC.dir.}}=N_{A}\mathcal{J}_{\mathrm{coh.dir.}}/x_{a}$.
Those for resolved contributions can be derived from the case of direct photoproduction processes by $\mathcal{J}/x_{b}$.
And for fragmentation processes, the Jacobian determinant can be presented as
\begin{eqnarray}\label{Jac.fragt}
\mathcal{J}=\frac{\hat{s}+Q^{2}}{\cosh(y_{r})\sqrt{\hat{s}}}\;.
\end{eqnarray}

We give the kinematical limitations for $Q^{2}$ and $p_{T}$ distributions in Table~\ref{Kinematical.VBQ2},~\ref{Kinematical.VBpT}.
Those for $y_{r}$ distribution are the same as Table~\ref{Kinematical.VBpT}, but $p_{T}$ should be integrated out,
\begin{align}
&p_{T\ \mathrm{min}}=1,\displaybreak[0]\nonumber\\
&p_{T\ \mathrm{max}}=\frac{1}{2\cosh y_{r}}\sqrt{\left[\frac{(\hat{s}_{\mathrm{max}}-M^{2})^{2}}{\hat{s}_{\mathrm{max}}}-4\sinh^{2}y_{r}M^{2}\right]}.
\displaybreak[0]\nonumber\\\label{Inte.Vari.yr}
\end{align}
The kinematical limitations for fragmentation processes are the same as above, but $\hat{s}$ and $\hat{s}_{\gamma}$ should be replaced by
its lower limits: $\hat{s}_{\mathrm{min}}=\hat{s}_{\gamma \mathrm{min}}=4\cosh^{2}y_{r}p_{T}^{2}$.

\begin{table*}[htbp]
\renewcommand\arraystretch{2.0}
\centering
\caption{\label{Kinematical.VBQ2} The bounds of integration variables for $Q^{2}$ distribution.
The bounds of variables for OIC. are the same as coh., but the term $s/N_{A}$ should be replaced by $s_{NN}$ in the case of Pb-Pb collision.
$\hat{s}_{\mathrm{min}}=\hat{s}_{\gamma \mathrm{min}}=(M_{T \mathrm{min}}+p_{T \mathrm{min}})^{2}$ and $p_{T}^{2}=\hat{t}(\hat{s}\hat{u}+Q^{2}M^{2})/(\hat{s}+Q^{2})^{2}$.
}
\begin{tabular}{L{1cm}cm{0.7cm}cm{0.7cm}cm{0.7cm}c}
\hline
\hline
variables & coh.dir. & & UIC.dir. & & coh.res. & & UIC.res. \\
    \hline
    $z_{q \mathrm{min}}$    & \multicolumn{7}{c}{$(M^{2}+\hat{s})/2\hat{s}-\sqrt{(\hat{s}-M^{2})^{2}-4p_{T \mathrm{min}}^{2}\hat{s}}/2\hat{s}$} \\
    \hline
    $z_{q \mathrm{max}}$    & \multicolumn{7}{c}{$(M^{2}+\hat{s})/2\hat{s}+\sqrt{(\hat{s}-M^{2})^{2}-4p_{T \mathrm{min}}^{2}\hat{s}}/2\hat{s}$} \\
    \hline
    $\hat{t}_{\textrm{min}}$ & $(z_{q \mathrm{min}}-1)yx_{b}s/N_{A}$ & & $(z_{q \mathrm{min}}-1)yx_{a}x_{b}s_{NN}$ & & $(z_{q \mathrm{min}}'-1)z_{a}yx_{b}s/N_{A}$ &
    & $(z_{q \mathrm{min}}'-1)z_{a}yx_{a}x_{b}s_{NN}$ \\
    \hline
    $\hat{t}_{\mathrm{max}}$ & $(z_{q \mathrm{max}}-1)yx_{b}s/N_{A}$ & & $(z_{q \mathrm{max}}-1)yx_{a}x_{b}s_{NN}$ & & $(z_{q \mathrm{max}}'-1)z_{a}yx_{b}s/N_{A}$ &
    & $(z_{q \mathrm{max}}'-1)z_{a}yx_{a}x_{b}s_{NN}$ \\
    \hline
    $z_{a \mathrm{min}}$    &  $\backslash$  & &  $\backslash$  & &  $N_{A}\hat{s}_{\gamma \mathrm{min}}/yx_{b}s$  & &  $\hat{s}_{\gamma\mathrm{min}}
    /yx_{a}x_{b}s_{NN}$ \\
    \hline
    $z_{a \mathrm{max}}$    &  $\backslash$  & &  $\backslash$  & &  1  & &  1 \\
    \hline
    $x_{b \mathrm{min}}$    & $N_{A}(\hat{s}_{\mathrm{min}}+Q^{2})/ys$ & & $(\hat{s}_{\mathrm{min}}+Q^{2})/yx_{a}s_{NN}$ & & $N_{A}\hat{s}_{\gamma \mathrm{min}}/z_{a \mathrm{max}}ys$ & & $\hat{s}_{\gamma \mathrm{min}}/z_{a \mathrm{max}}yx_{a}s_{NN}$ \\
    \hline
    $x_{b \mathrm{max}}$    &  \multicolumn{7}{c}{1}  \\
    \hline
    $x_{a \mathrm{min}}$    &  $\backslash$  & &  $(\hat{s}_{\mathrm{min}}+Q^{2})/ys_{NN}$  & &  $\backslash$  & &  $\hat{s}_{\gamma \mathrm{min}}/z_{a \mathrm{max}}ys_{NN}$ \\
    \hline
    $x_{a \mathrm{max}}$    &  $\backslash$  & &  1  & &  $\backslash$  & &  1 \\
    \hline
    $y_{\mathrm{min}}$       & \multicolumn{3}{c}{$N_{A}(\hat{s}_{\mathrm{min}}+Q^{2})/s$}  & &  \multicolumn{3}{c}{$N_{A}\hat{s}_{\gamma \mathrm{min}}/z_{a \mathrm{max}}s$} \\
    \hline
    $y_{\mathrm{max}}$       & \multicolumn{7}{c}{$\left[\sqrt{Q^{2}(4m_{\alpha}^{2}+Q^{2})(m_{\alpha}^{2}-s/N_{A})^{2}}+(m_{\alpha}^{2}-s/N_{A})Q^{2}\right]N_{A}/2m_{\alpha}^{2}s$} \\
\hline
\hline
\end{tabular}
\end{table*}

\begin{table*}[htbp]
\renewcommand\arraystretch{2.0}
\centering
\caption{\label{Kinematical.VBpT} Same as Table \ref{Kinematical.VBQ2} but for $p_{T}$ distribution. $x_{1}=N_{A}\hat{s}/s$, $z_{a \mathrm{max}}=1/(1+Q^{2}/4p_{T}^{2})$ \cite{Rossi:1983xz}.
The bounds of $y$ are the same as Table~\ref{Kinematical.VBQ2}, we are not list it here.
}
\begin{tabular}{C{2.5cm}cm{0.7cm}cm{0.7cm}cm{0.7cm}c}
\hline
\hline
variables & coh.dir. & & UIC.dir. & & coh.res. & & UIC.res. \\
    \hline
    $Q^{2}_{\mathrm{min}}~[\mathrm{GeV}^{2}]$   &  \multicolumn{7}{c}{$x^{2}_{1}m^{2}_{\alpha}/(1-x_{1})$} \\
    \hline
    $Q^{2}_{\mathrm{max}}~[\mathrm{GeV}^{2}]$   &  $N_{A}^{-2/3}0.027$  & &  $4p_{T}^{2}$  & &  $N_{A}^{-2/3}0.027$ & &  $4p_{T}^{2}$ \\
    \hline
    $|y_{r \mathrm{max}}|$  & \multicolumn{7}{c}{$\ln\left[\left(\hat{s}_{\mathrm{max}}+M^{2}+\sqrt{(\hat{s}_{\mathrm{max}}-M^{2})^{2}
    -4p^{2}_{T}\hat{s}_{\mathrm{max}}}\right)/\left(\hat{s}_{\mathrm{max}}+M^{2}-\sqrt{(\hat{s}_{\mathrm{max}}
    -M^{2})^{2}-4p^{2}_{T}\hat{s}_{\mathrm{max}}}\right)\right]/2$} \\
    \hline
    $x_{b \mathrm{min}}$    &  $\backslash$  & &  $\backslash$  & &  $N_{A}\hat{s}_{\gamma}/z_{a \mathrm{max}}ys$  & &  $\hat{s}_{\gamma}/z_{a \mathrm{max}}yx_{a}s_{NN}$ \\
    \hline
    $x_{b \mathrm{max}}$    &  $\backslash$  & &  $\backslash$  & &  1  & &  1 \\
    \hline
    $x_{a \mathrm{min}}$    &  $\backslash$  & &  $(\hat{s}+Q^{2})/ys_{NN}$  & &  $\backslash$  & &  $\hat{s}_{\gamma}/z_{a \mathrm{max}}ys_{NN}$ \\
    \hline
    $x_{a \mathrm{max}}$    &  $\backslash$  & &  1  & &  $\backslash$  & &  1 \\
\hline
\hline
\end{tabular}
\end{table*}

\end{document}